\documentclass[graybox]{svmult}

\usepackage{aas_macros}
\usepackage{type1cm}        
\usepackage{makeidx}         
\usepackage{graphicx}        
\usepackage{multicol}        
\usepackage[bottom]{footmisc}

\usepackage{xspace}
\usepackage{newtxtext}       %
\usepackage{amsmath}
\usepackage{newtxmath}       
\usepackage{comment}
\usepackage{subcaption}
\makeindex             

\def \inte {{\em INTEGRAL\xspace}}
\def \ibis  {{\em IBIS/ISGRI\xspace}}
\def \swift {{\em Swift\xspace}}
\def \chandra {{\em Chandra\xspace}}
\def \xmm {{\em XMM-Newton\xspace}}
\def \rxte {{\em RXTE\xspace}}
\def \suzaku {{\em Suzaku\xspace}}
\def \saxj{{\rm SAX~J1808.4$-$3658\xspace}}

\def \igrsev{{\rm IGR~J17591$-$2342\xspace}}

\def \swiftxrt{{\em Swift/XRT\xspace}}
\def \swiftbat{{\em Swift/BAT\xspace}}
\def \nustar{{\em NuSTAR\xspace}}
\def \nicer{{\em NICER\xspace}}
\def \maxigsc{{\em MAXI/GSC\xspace}}

\DeclareUnicodeCharacter{2212}{-}

\makeatletter
\newcommand*{\rom}[1]{\expandafter\@slowromancap\romannumeral #1@}
\makeatother

\begin{document}

\title*{Accretion Powered X-ray Millisecond Pulsars}
\author{Tiziana Di Salvo and Andrea Sanna}
\institute{Universit\`a degli Studi di Palermo, Dipartimento di Fisica e Chimica - Emilio Segrè, via Archirafi 36, 90123 Palermo, Italy, \email{tiziana.disalvo@unipa.it}
\and Universit\`a degli Studi di Cagliari, Dipartimento di Fisica, SP Monserrato-Sestu km 0.7, 09042 Monserrato, Italy \email{andrea.sanna@dsf.unica.it}}
%
%
\maketitle

\abstract*{}

\abstract{}
Neutron Stars are among the most exotic objects in the Universe. A neutron star, with a mass of 1.4 -- 2$\, M_\odot$ within a radius of about 10 -- 15 km, is the most compact stable configuration of matter in which degeneracy pressure can still balance gravity, since further compression would lead to gravitational collapse and formation of a black hole. As gravity is extreme, rotation is extreme: neutron stars are the fastest rotating stars known, with periods as short as a millisecond. The presence of a magnetic field not aligned with the rotation axis of the star is the origin of pulsating emission from these sources, which for this reason are dubbed pulsars. The discovery in 1998 of the first Accreting Millisecond X-ray Pulsar, started an exciting season of continuing discoveries. In the last 20 years, thanks to the extraordinary performance of astronomical detectors in the radio, optical, X-ray, and Gamma-ray bands, astrophysicists had the opportunity to thoroughly investigate the so-called Recycling Scenario: the evolutionary path leading to the formation of a Millisecond-spinning Pulsar. 
In this chapter we review the general properties of Accreting Millisecond X-ray Pulsars, which provide the first evidence that neutron stars are spun up to millisecond periods by accretion of matter and angular momentum from a (low-mass) companion star. 
We describe the general characteristics of this class of systems with particular attention to their spin and orbital parameters, their short-term and long-term evolution, as well as the information that can be drawn from their X-ray spectra.

\section{How to spin up a Neutron Star: the recycling scenario}

Millisecond Pulsars (hereafter MSPs) are fast-spinning Neutron Stars (NS) with periods shorter than 30 ms, and hence spin frequency higher than 30 Hz. Now we know that the vast majority of these fast-spinning NS are in binary systems with a low-mass ($< 1\, M_\odot$) companion star and possess a relatively weak magnetic field (less than $10^8$ -- $10^9$ Gauss). Moreover, a large amount of these systems are found in Globular Clusters (old clusters of stars). It was soon realised that these NS must belong to old systems in order to have the time for the magnetic field to decay from the large strength in young NS (usually above $10^{12}$ Gauss) to their present, much lower strength. It was therefore proposed that old NS are spun up to millisecond periods by the accretion of matter and angular momentum during a Low Mass X-ray Binary (hereafter LMXB) phase; this is the so-called {\it recycling scenario} (see e.g. \cite{Bhattacharya1991}. Once a {\it recycled} NS reaches a high spin frequency, even if its magnetic field has decayed, the rotation-powered emission mechanism (which depends on the fourth power of the spin frequency, see below) can be re-activated; hence, at the end of the accretion phase, when the companion star has lost its atmosphere and/or has detached from its Roche lobe, the NS should be visible as a rotation-powered MSP.

\subsection{Evolution of rotation-powered Neutron Stars in the $P - \dot P$ diagram}

According to the recycling scenario, a newly born NS should have on average a relatively slow spin period (above few tens of milliseconds) and a relatively strong magnetic field; an example is given by the Crab pulsar, a 33 ms isolated pulsar with a magnetic field of $\sim 4 \times 10^{12}$ Gauss discovered in 1968 at the center of a young, $\sim 1000$-years old, supernova remnant called the Crab Nebula. The magnetic field, rotating at the spin period of the NS, behaves as a rotating magnetic dipole which emits radiation according to the Larmor formula (e.g. \cite{Jackson}):  
\begin{equation}
\label{Larmor}
P_{\rm rad} = \frac{2}{3} \frac{(\ddot{\mu}_\bot)^2}{c^3} = 
\frac{2}{3}\frac{\mu_\bot^2 \Omega^4}{c^3} = \frac{2}{3c^3}( B R^3 \sin \alpha)^2 \biggl( \frac{2\pi}{P} \biggr)^4~,
\end{equation}
where $\mu_\bot = B R^3 \sin \alpha$ is the component of the magnetic dipole moment perpendicular to the rotation axis, $B$ and $R$ are the surface magnetic field and the NS radius, respectively, $\alpha$ is the angle between the rotation axis and the magnetic dipole axis, $\Omega$ is the spin angular frequency of the NS and $P$ its spin period. 
In this case the (pulsed) emission is usually visible in the radio (and often in the gamma-ray) band; it is mainly due to synchrotron emission of charge currents, formed by electrons and positrons extracted from the NS surface by the intense Lorentz force due to the magnetic field and the fast rotation, moving along curved open magnetic field lines \cite{Jackson}. Because of this emission, the NS looses rotational energy and slows down, according to the relation (see e.g. \cite{Spitkovsky2006}):
\begin{equation}
\label{spin-down}
\dot E = \frac{d}{dt} \left(\frac{1}{2} I \Omega^2\right) = I \Omega \dot \Omega \simeq -\frac{\mu^2 \Omega^4}{c^3}  (1+\sin^2 \alpha),
\end{equation}
where $I \propto M R^2$ is the moment of inertia of the NS, $M$ its mass, and $\mu = B_0 R^3/2$ is the magnetic moment, where $B_0$ is the magnetic field strength at the pole. Solving this equation for $B_0$ and inserting typical values of $I \simeq 10^{45}$ g cm$^2$, $R\simeq10^6$ cm and $\alpha = 90^\circ$, gives: 
\begin{equation}
\label{B0}
B_0 \sim 6 \times 10^{19} (P \dot P)^{1/2}\, \rm Gauss,
\end{equation}
which allows to relate the magnetic field strength with the spin period and its derivative, and hence to give an estimate of the magnetic field once the spin period and its derivative are measured. In the hypothesis that the magnetic field does not change significantly with time, from eq.~(\ref{spin-down}) we can estimate the pulsar characteristic age $\tau = P / (2 \dot P)$, defined as the timescale necessary to bring the pulsar from its initial period $P_0$ to the current period $P$ at the observed spin-down rate by assuming $P_0<<P$. For instance the characteristic age of the Crab pulsar comes out to be $\sim 2.5$ kyr. 

In the meantime, the NS magnetic field rapidly decays due to mechanisms not fully understood yet (probably ohmic dissipation, see e.g. \cite{Tauris2001} for a review, perhaps with a contribution due to accretion of matter in the subsequent LMXB phase, see e.g \cite{Cumming2001}), and causes the pulsar to move down almost vertically in a $P-\dot P$ (or magnetic field strength vs.\ P) diagram. Below the so-called death line, the NS enters the {\textit graveyard} where the rotation-powered pulsar switches off because not enough spin-down power is available to feed the emission mechanism. At this point, if the NS is in a binary system with a low-mass star (less than $1\, M_\odot$), the latter may be able to fill its Roche-lobe due to nuclear evolution and/or losses of the orbital angular momentum caused by Magnetic Braking (MB) of the companion star and/or Gravitational Radiation (GR). This scenario envisages mass-transfer phases in which the system will start emitting in X-rays and will be observed as a LMXB. At the end of the LMXB phase, the NS, spun-up to millisecond periods by the accretion of matter and its angular momentum, can exit the graveyard and be observed as a rotation-powered MSP.

\subsection{Low-Mass X-ray Binaries and accretion onto a Neutron Star}

LMXBs are (Gyr-old) binary systems in which a low-mass star transfers matter onto a compact object via Roche lobe overflow. Matter passing through the inner Lagrangian point has a large specific angular momentum (because of the rotation of the system around its center of mass) that inhibits radial accretion. Matter starts to spiral-in around the compact obiect creating a structure usually defined as \textit{accretion disk}, in which internal torques transfer angular momentum outwards, allowing the accreting plasma to slowly approach the compact object. The mechanical energy of matter is partially dissipated in the disk which emits a blackbody-like spectrum with a temperature increasing towards the center. If the compact object is a NS, the rest of the mechanical energy of the matter is released when the accreting matter reaches the surface of the NS, giving a total luminosity of $\sim G M_{NS} \dot M / R_{NS}$, where $\dot M$ is the mass accretion rate onto the NS. This implies an efficiency in the conversion of rest mass energy to luminosity of $\eta = G M_{NS} / (c^2 R_{NS}) \sim 0.21$ for a $1.4\, M_\odot$ NS with 10 km radius. The mass accretion rate onto the NS is limited by radiation pressure which can be high enough to balance the gravitational force towards the NS. This happens at the so-called Eddington limit; in the hypothesis of stationary and spherical accretion, the maximum luminosity of the system is given by: $L_{Edd} \simeq 1.3 \times 10^{38}\, M/M_\odot$ erg s$^{-1}$.
For a NS with $M_{NS} = 1.4\, M_\odot$, the Eddington limit is given by $L_{Edd} \simeq 2.5 \times 10^{38}$ erg s$^{-1}$ (appropriate for helium-rich material and a moderate, $z=1.2$, gravitational redshift correction factor, \cite{vanParadijs1994}), corresponding to a mass accretion rate of $\dot M_{Edd} \sim 1.3 \times 10^{18} g s^{-1} \sim 2 \times 10^{-8}\, M_\odot\, yr^{-1}$. A the Eddington luminosity, the  blackbody emission from the NS surface will reach a temperature of about 20 million K, corresponding to $\sim 2$ keV in photon energy, and implying that the emission from the innermost region of the system will be mainly in the X-ray band. At such high luminosity, strong outflows of matter can rise driven by the strong radiation pressure, and the inner part of the disk may inflate and form a geometrically thick and optically thick disk, also known as {\it thick disk}. 
In general, besides the blackbody-like components produced by the accretion disk and the NS surface, the X-ray spectrum of a LMXB is often complicated by the presence of a hot electron corona in the central part of the system, which up-scatters soft photons coming from the disk and/or the NS surface, producing hard Comptonization spectra.

In the case of a magnetised NS, the accretion flow towards the NS can be halted by the magnetosphere depending on the magnetic field strength. For a dipolar magnetic field, the magnetic energy ($B^2/8\pi$) increases at small radii and can overcome the Kinetic energy ($\rho v^2/2$) of the (free-falling) in-falling matter at the magnetospheric radius (which delimits the NS magnetosphere); inside this radius (charged) particles are forces to flow along the magnetic field lines and will be accreted at the NS polar caps. For spherical accretion, this radius is called Alfv\'en radius and is given by:
\begin{equation}
\label{Alfven}
R_A = \left(\frac{\mu^4}{2 G M_{NS} \dot M^2}\right)^{1/7} \sim 3.7 \times 10^6 \mu_{26}^{4/7} \dot M_{-10}^{-2/7} (M/M_\odot)^{-1/7}\, \rm cm
\end{equation}
where $\mu_{26}$ is the magnetic moment ($B R^3$) in units of $10^{26}$ Gauss cm$^3$ and $\dot M_{-10}$ is the mass accretion rate in units of $10^{-10}\, M_\odot$ yr$^{-1}$. It is easy to deduce that, in order to have a magnetospheric radius larger than the NS radius for a mass accretion rate of about $10\%$ of the Eddington limit, the magnetic field should be higher than $10^8$ Gauss. In this case, under the hypothesis of magnetic axis not aligned with the NS spin axis, accretion onto the polar caps can produce a lighthouse signal visible as X-ray (accretion-powered) pulsations, which give us a direct measure of the NS spin period.
For disk-fed accretion flows, the ram pressure of matter is concentrated in the disk plane, allowing it to penetrate further the NS magnetosphere and reducing the magnetospheric radius, $r_m$, with respect to the Alfv\'en radius, by a factor $\sim$ 0.3 -- 0.5 (see e.g. \cite{Ghosh1991,Burderi1998}).

The interaction of the accretion flow with the NS magnetosphere allows an exchange of angular momentum between the accreting matter and the NS which results in a spin-up or spin-down of the NS (see e.g. \cite{Ghosh1979} for a first study of the torques exerted by the accreting matter onto the NS). Assuming that the inner accretion disk is truncated at the magnetospheric radius and defining the co-rotation radius $r_{CO}$ as the radius at which the Keplerian angular velocity of the disk, $\omega_K(r)$, matches the NS angular velocity $\Omega_0$, that is $r_{CO} = (G M_{NS} / \Omega^2_0)^{1/3} \sim 2.8 \times 10^6 (M / M_\odot)^{1/3} P_{ms}^{2/3}$ cm (where $P_{ms}$ is the NS spin period in millisecond), we can envisage the following three possibilities: 

i) $r_m < r_{CO}$, i.e. the inner disk rotates faster than the magnetosphere and exerts a positive torque spinning-up the NS. In this case the spin-up torque is given at zero order by: $I \Omega \dot \Omega = \dot{M} (G M_{NS} r_m)^{1/2}$, i.e. by the mass accretion rate onto the NS times the specific angular momentum at the magnetospheric radius (the latter also has a weak dependence on the mass accretion rate, as $\dot M^{-2/7}$). According to \cite{Ghosh1979}, the rate of change of the NS period is:
\begin{equation}
\frac{\dot P}{P} = -3 \times 10^{-8} f \frac{P} {1 ms} \left(\frac{L_X}{10^{37} erg s^{-1}}\right)^{6/7} yr^{-1},
\end{equation}
where the dimensionless parameter $f$ is expected to be of the order of unity. This demonstrates that in the LMXB phase the NS can be efficiently spun-up by accretion torques within its lifetime. Moreover, it can be shown that, for a slow-rotating low-magnetic field NS, it is enough to accrete no more than $\sim$ 0.1 -- 0.2 $M_\odot$ to spin-up the NS to millisecond periods (see e.g. \cite{Burderi1999}), unless the mass transfer is highly not conservative. Hence, for a NS it is in principle possible to reach mass-shedding spin periods before the gravitational collapse into a black hole.

ii) $r_m > r_{CO}$, i.e.\ the Keplerian velocity at the inner accretion disk is lower than the angular velocity of the NS and this causes a spin-down torque onto the NS. Indeed, in this case, the centrifugal barrier should prevent matter to penetrate the magnetosphere, giving the so-called {\it propeller effect}. However, magneto-hydrodynamic simulations \cite{Romanova2005} suggest that this is true only when the magnetospheric radius is very large when compared to $r_{CO}$ ($r_m >> r_{CO}$, strong propeller regime), otherwise matter can still (at least in part) penetrate the magnetosphere and accrete onto the NS (weak propeller regime), allowing the possibility to observe spin-down during (low-rate) accretion phases. The latter may also be favoured by some threading of the magnetic field lines by the accretion disk beyond the co-rotation radius (see e.g. \cite{Wang87, Rappaport2004, Kluzniak2007}, and references therein) when magnetic field lines are not completely shielded by current sheets at the magnetospheric radius. In this case, threading of the magnetic filed can result in both a spin-up (due to threading inside $r_{CO}$) and a spin-down (due to threading outside $r_{CO}$), and the balance of the two, plus the material torque, gives the net torque exerted onto the NS. The possibility to have a spin-down of the NS during accretion phases for fast rotators has been studied by \cite{Rappaport2004, Kluzniak2007}; these authors argue that the accretion disk structure around a fast pulsar will adjust itself so that the inner edge of the disk, also known as the truncation radius, will remain fixed near $r_{CO}$ while accretion will continue. In this case, the net torque onto the NS is given by the accretion torque of matter captured at the co-rotation radius decreased by a spin-down torque due to the magnetic field drag on the accretion disk, which, at a first order of approximation, can be expressed as $\mu^2 / (9 r_{CO}^3)$, resulting in the net torque:
\begin{equation}
\label{torque}
\tau_{NS}= 2\pi I \dot{\nu}_{NS} = \dot M (G M_{NS} r_{CO})^{1/2} - \frac{\mu^2}{9 r_{CO}^3}.
\end{equation}
iii) $r_m \sim r_{CO}$, in this case matter loaded by the magnetic field will have the same angular velocity of the NS and no net torque is expected onto the NS, meaning that the NS will be at the spin equilibrium. In this case, the equilibrium period is given by:
\begin{equation}
\label{Peq}
P_{eq} = 0.5 \mu_{26}^{6/7} L_{37}^{-3/7} R_6^{-3/7} (M/M_\odot)^{-2/7}\, \rm ms,
\end{equation}
where $R_6$ is the NS radius in units of $10^6$ cm. This means that a NS can reach in principle a spin equilibrium period shorter than a millisecond (see Sec.~3 on Chapter~7 for similar discussions based on the MSPs changes of state). However, the maximum spin frequency that a NS can attain also depends on its Equation of State (EoS, see e.g.\ \cite{Ozel2016} for a review), which sets the mass-shedding spin limit depending on the mass-radius relation. The fastest-spinning NS known to date is PSR J1748-2446ad, a rotation-powered pulsar spinning at 716 Hz \cite{Hessels2006}. This spin frequency is, however, not high enough to put strong constraints onto the NS EoS and it is not clear yet whether other mechanisms (e.g.\ GR emission, a relatively large magnetic field, observational bias, and so on) can be responsible of the lack of ultra-fast spinning NS (see e.g. \cite{Burderi2001,Patruno2017b,Haskell2018}).

\section{The discovery of Accreting X-ray Millisecond Pulsars: the missing link in the recycling scenario}

From what is discussed above, it appears clear that NS can be spun-up to millisecond periods or below, depending on the constraints imposed by the EoS of ultra-dense matter, during the accretion phase in a LMXB. However, till 1998, no LMXB was found to show any coherent pulsation at such low periods. The fact that the vast majority of LMXBs does not show coherent pulsations is still a fascinating enigma. Several possible explanations have been invoked to interpret this fact, but none of them is fully satisfactory (see also \cite{Patruno2012} and references therein, for further discussion of this issue). Thanks to the large effective area ($\sim 6500\, cm^2$) and good timing capabilities ($\sim 1\, \mu s$) of the NASA satellite Rossi X-ray Timing Explorer (\rxte{}), in 1998 \cite{Wijnands1998} discovered the first LMXB, \saxj, to show coherent pulsations (at about $401$ Hz). This was the first direct confirmation of the recycling scenario, since it demonstrated that LMXBs could indeed host a fast-spinning NS. Doppler effects visible in the spin period of the pulsar revealed the $\sim 2$ h orbital period of the system  \cite{Chakrabarty1998}. 
The final confirmation of the recycling scenario arrived only in 2013, when \cite{Papitto2013b} discovered a transient system, IGR~J18245-2452, showing accretion-powered pulsations during the X-ray outburst and rotation-powered radio pulsations during X-ray quiescence. This source, which is one of the members of the so-called {\it transitional} MSP class, is the direct evidence of the fact that, when accretion stops, the rotation-powered pulsar mechanism should resume on a short timescale.

\saxj, first discovered in 1996 by the Wide Field Camera (WFC) on board the X-ray satellite BeppoSAX, is a transient system, which spends most of the time in quiescence (with X-ray luminosity around few $10^{31}$ erg s$^{-1}$, \cite{Campana2004}) and shows month-long X-ray outbursts every $\sim3$ yr, during which it reaches an X-ray luminosity in the range $10^{36}$ -- $10^{37}$ erg s$^{-1}$. Now we know about two dozens of these systems, belonging to the class of Accreting Millisecond X-ray Pulsars (hereafter AMXPs), most of them discovered by \rxte{} and the ESA satellite \xmm{}, and more recently by \nustar\ and \nicer{}.
All of them are transient systems, although with very different transient behaviour (see Table 2 in \cite{2019A&A...627A.125M} for an overview). X-ray outbursts usually last from few days to less than three months. Most of the AMXPs have shown just one outburst since their discovery, while a few sources show recurrent outbursts. The shortest outburst recurrence time is about a month, registered for the globular cluster source NGC 6440 X-2, with an outburst duration of less than $4-5$ days, whereas the longest outburst has been observed from HETE J1900.1-2455, and has lasted for about 10 years (up to late 2015 when the source returned to quiescence, \cite{Degenaar2017b}).

Another peculiar behaviour is the intermittency of the pulsations, which is important because the understanding of this phenomenon could give insights on the lack of X-ray pulsations in the large majority of NS LMXBs. This phenomenon was observed for the first time in the AMXP HETE J1900.1-2455, which went into X-ray outburst in 2005 and showed X-ray pulsations at 377 Hz. However, after the first 20 days of the outburst, pulsations became intermittent for about 2.5 yr, and then disappeared with very stringent upper limits on the pulsed fraction ($< 0.07\%$; \cite{Patruno2012a}). The most peculiar behaviour was observed from Aql X-1, a transient LMXB showing regular outbursts more or less every 0.5 -- 1 year (see e.g. \cite{Campana2013}); it showed coherent pulsations in only one 150-s data segment out of a total exposure time of $\sim1.5$ Ms from more than 10 years of \rxte{} monitoring \cite{Casella2008}. Another AMXP showing intermittency of pulsations is SAX J1748.9-2021, where pulsations were detected sporadically in several data segments and in three out of four outbursts observed by the source (see e.g. \cite{Patruno2009}). Note that these AMXPs may have a long-term average mass accretion rate higher than the other AMXPs. To explain this {\it intermittent} behaviour, it has been proposed that the accreting matter could screen the NS magnetic field, weakening it by orders of magnitude on a few hundred days timescale, hampering the possibility to effectively channel the accretion flow towards the NS polar caps (e.g. \cite{Patruno2012a}). However, it is not clear yet whether this hypothesis can explain all the phenomenology observed in AMXPs, and more observations and theoretical efforts are needed to reach a satisfactory explanation of this puzzling behaviour.

In Table~\ref{tab:lmxbs} we show the main properties of the AMXPs known to date. The following sections will be dedicated to the description of the main results obtained to date on the spectral and timing properties of this class of sources. 

\begin{table}
\caption{Accreting X-ray Pulsars in Low Mass X-ray Binaries.}
\scriptsize
\begin{center}
\begin{tabular}{lllllll}
\hline
\hline
Source & $\nu_{s}/P$ & $P_{\rm orb}$ & $f_{x}$  & $M_{c,min}$  & Companion  & Ref.\\
 & (Hz)/(ms) & (hr) & ($M_{\odot}$) & ($M_{\odot}$) &   Type & \\
\hline
\textbf{Accreting Millisecond X-ray Pulsars}\\
\hline
Aql X-1     & 550 (1.8) &  18.95 & $1.4\times 10^{-2}$ & 0.56 & MS & \cite{Casella2008,MataSanchez2017}\\
IGR J17591-2342 & 527 (1.9) &  8.80 & $1.5\times 10^{-2}$ & 0.37 & MS & \cite{Sanna2018c}\\
Swift J1749.4-2807 & 518 (1.9) & 8.82 & $5.5\times 10^{-2}$ & 0.59 & MS & \cite{Altamirano2011,DAvanzo2011}\\
SAX J1748.9-2021  & 442 (2.3) & 8.77 &  $4.8\times 10^{-4}$ & 0.1  & MS& \cite{Altamirano2008,Cadelano2017}\\
IGR J17498-2921 & 401 (2.5) & 3.84 & $2.0\times10^{-3}$ & 0.17 & MS & \cite{Papitto2011b}\\
XTE J1814-338  & 314  (3.2) & 4.27 & $2.0\times 10^{-3}$ & 0.17 & MS  & \cite{Markwardt2003,Wang2017}\\
IGR J1824-2453 & 254 (3.9) &  11.03 & $2.3\times 10^{-3}$ & 0.17  & MS & \cite{Papitto2013b}\\
IGR J17511-3057 & 245 (4.1) &  3.47 & $1.1\times 10^{-3}$ & 0.13  & MS & \cite{Papitto2010}\\
\hline
IGR J00291+5934    & 599  (1.7) & 2.46 & $2.8\times 10^{-5}$ & 0.039  &  BD & \cite{Galloway2005}\\
IGR J17379-3747    & 468  (2.1) & 1.88 & $8\times 10^{-5}$ & 0.056  &  BD & \cite{Sanna2018c}\\
SAX J1808.4-3658 & 401 (2.5) &  2.01 & $3.8\times 10^{-5}$ & 0.043  & BD &  \cite{Wijnands1998,Wang2013}\\
HETE J1900.1-2455& 377  (2.7) &   1.39 & $2.0\times 10^{-6}$ & 0.016  & BD  & \cite{Kaaret2006,Elebert2008}\\
\hline
XTE J1751-305 & 435 (2.3) &  0.71 & $1.3\times 10^{-6}$ & 0.014  & He WD &   \cite{Markwardt2002,DAvanzo2009}\\
MAXI J0911-655 & 340 (2.9) & 0.74 & $6.2\times 10^{-6}$ & 0.024  & He WD? &   \cite{Sanna2017a}\\
NGC6440 X-2 & 206 (4.8) & 0.95 & $1.6\times 10^{-7}$ & 0.0067 & He WD & \cite{Altamirano2010}\\
Swift J1756.9-2508  & 182  (5.5) &  0.91 &  $1.6\times 10^{-7}$ & 0.007  & He WD &  \cite{Krimm2007}\\
IGR J17062-6143  & 164  (6.1) &  0.63 &  $9.1\times 10^{-8}$ & 0.006  & He WD? &  \cite{Strohmayer2017}\\
IGR J16597-3704 & 105 (9.5) & 0.77 & 1.2$\times 10^{-7}$ & 0.006 & He WD & \cite{Sanna2018a}\\
\hline
XTE J0929-314  & 185  (5.4) & 0.73 & $2.9\times 10^{-7}$ & 0.0083  & C/O WD  & \cite{Galloway2002,Giles2005}\\
XTE J1807-294  & 190  (5.3) & 0.67 & $1.5\times 10^{-7}$ & 0.0066  & C/O WD  & \cite{Campana2003,DAvanzo2009} \\
\hline
\hline
\end{tabular}\\
\end{center}
$\nu_{s}$ is the spin frequency, $P_{b}$ the orbital period, $f_{x}$
is the X-ray mass function, $M_{c,min}$ is the minimum companion mass, calculated for an inclination $\sin i =1$ of the
binary system and for an assumed NS mass of 1.4 M$_\odot$. 
The companion types are: WD = White Dwarf, BD= Brown Dwarf, MS = Main Sequence, He Core = Helium Star.\newline
$^{b}$ Binary with parameters that are still compatible with an intermediate/high mass donor.\newline
Adapted and updated from \cite{Patruno2012a}.
\label{tab:lmxbs}
\end{table}

\section{Timing and Spectral Properties of AMXPs}

\subsection{Spectral properties}

\begin{figure}
\begin{subfigure}{.5\textwidth}
  \centering
  \includegraphics[angle=-90,width=1.1\linewidth]{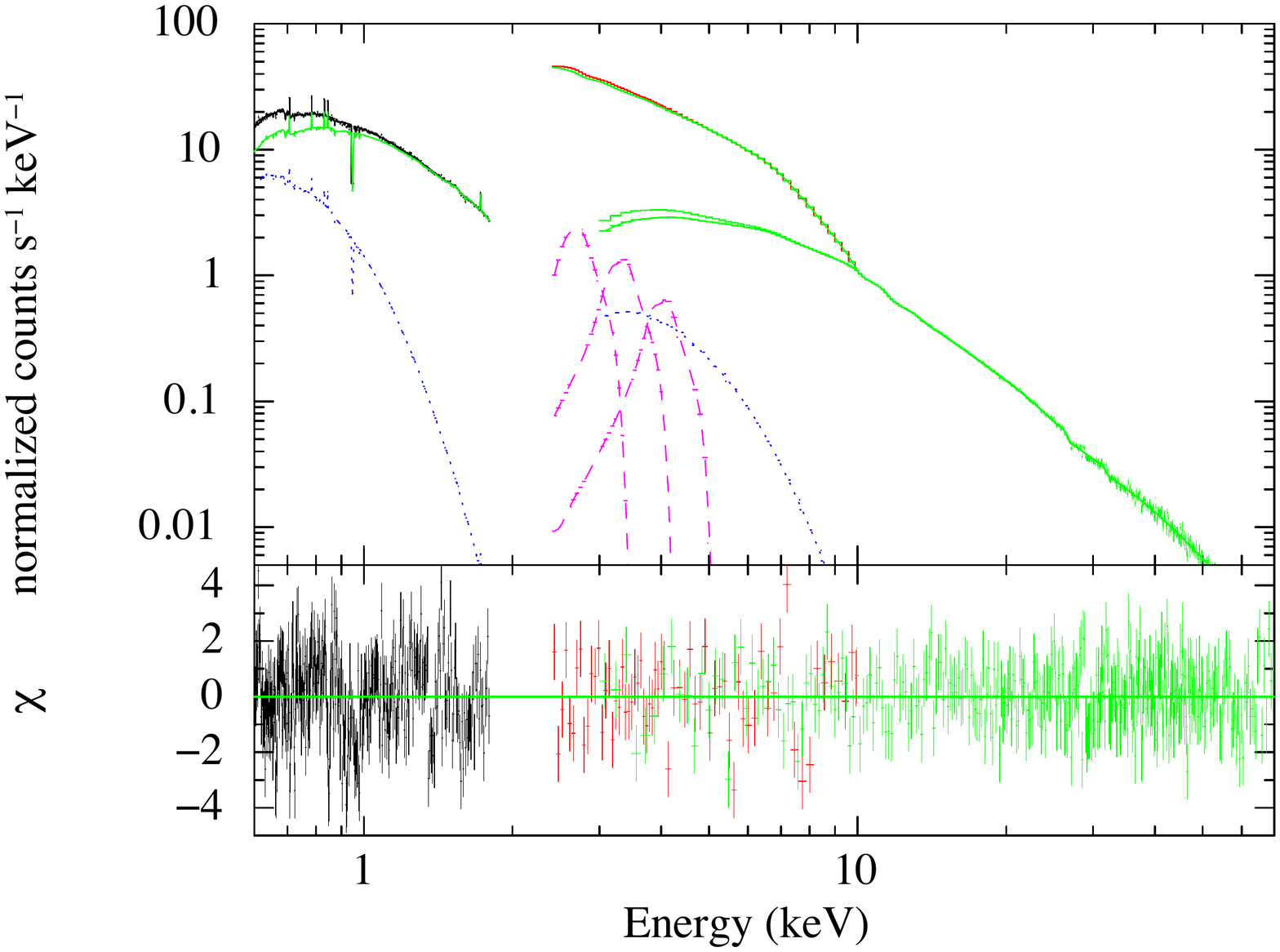}
  \label{fig:sfig1}
\end{subfigure}%
\begin{subfigure}{.5\textwidth}
  \centering
  \includegraphics[width=0.95\linewidth]{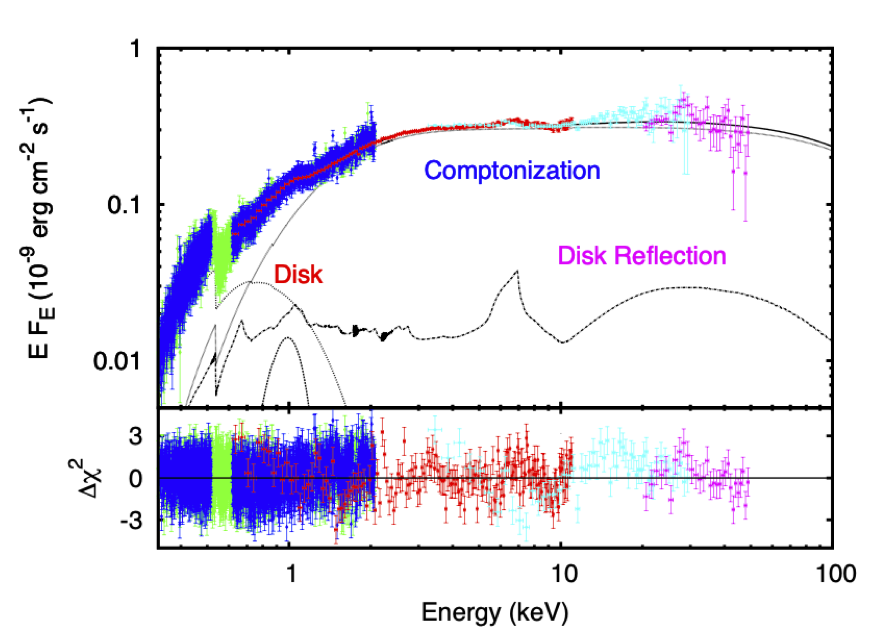}
  \label{fig:sfig2}
\end{subfigure}
\caption{\textit{Left Panel:} Broad-band spectrum of the AMXP \saxj{} during its 2015 outburst as observed by \xmm{} (black and red points) and \nustar{} (green points). The model includes a blackbody component, the {\it relxillCP} component, which includes the Comptonization continuum and the smeared reflection component, and three low-energy features modelled with {\it disklines}. \textit{Right Panel:} Broad-band spectrum of the AMXP HETE J1900.1-2455 in outburst as observed by \xmm{} (blue, green and red points) and \rxte{} (cyan and magenta points). The single spectral components used to fit the X-ray spectrum are indicated in the figure. The Gaussian feature at 0.98 keV may be identified with Fe L$\alpha$ or He-like Ne K$\alpha$ transition. [Figures from \cite{DiSalvo2019,Papitto2013a}]}
\label{fig:spec}
\end{figure}

In the vast majority of the AMXPs, the X-ray luminosity during outburst remains below $10\%$ the Eddington luminosity, and the spectra do not show transitions between hard and soft spectral states, as it usually happens for non-pulsating LMXBs (harbouring a NS or a black hole, see e.g. \cite{Done2007}). For this reason, AMXPs are often referred to as hard X-ray transients. Hence, their spectra are quite similar to the spectra usually observed for NS LMXBs in the hard state, with little spectral evolution during the X-ray outburst. In particular, the X-ray continuum is composed of one or two blackbody-like components and an unsaturated Comptonization component, usually with cutoff energies (corresponding to the electron temperature of the Comptonizing cloud, often called {\it corona}) of tens of keV \cite{Gierlinski2002,Gierlinski2005,Poutanen2006}. In this case, the presence of a {\it reflection} of the hard Comptonization photons off the cold accretion disk is expected. This reflection component usually contains discrete features, such as the florescence iron line at 6.4 -- 6.7 keV (depending on the iron ionization state), which are smeared by Doppler and relativistic effects due to the large velocity of matter in the inner accretion disk (see the right panel of Fig.~\ref{fig:spec} where these spectral components are indicated). The precise modelling of these features may give information about some important physical parameters as ionization state of matter in the inner accretion disk, the inclination of the system with respect to the line of sight, the radius at which the inner accretion disk is truncated, the outer radius of the emitting region in the disk, and the index of the emissivity law in the disk, which is $\propto r^{index}$. 

The inner disk radius is an important parameter since it may be useful to obtain an estimate of the magnetospheric radius, which can be compared to the co-rotation radius of the source in order to test the accretion torque paradigm described above. Together with discrete features (emission lines and absorption edges), the hard photons impinging the disk are scattered by Thomson/(direct) Compton effect, generating a continuum spectrum (with a shape similar to the primary Comptonization spectrum) which is usually evident as an excess of emission between 10 and 30 keV named Compton hump. The spectral shape of this continuum is also sensitive to the inclination angle of the system and the ionization state of matter in the disk. The latter is measured through the parameter $\xi = L_X/(n_e r^2)$, where $L_X$ is the bolometric X-ray luminosity of the ionizing continuum, $n_e$ is the electron density in the disk atmosphere and $r$ is the distance of the disk from the center of the system. For high values of the ionization parameter $\xi$, photoelectric absorption of soft X-rays in the disk is suppressed and this results in a strong reflection continuum, which increases at soft X-ray energies instead of decreasing.

Most, but not all, of the AMXPs have been observed with moderate/high resolution instruments in order to perform a broad-band spectral analysis and look for reflection features. In the following we describe the main spectral results obtained for the available sample of AMXPs, with particular attention to reflection features and the constraints that can be inferred on the inner accretion flow. 
Indeed, a relatively small inner disk radius is implied for most of the AMXPs for which a spectral analysis has been performed and a broad iron line has been detected in moderately high resolution spectra. 
The AMXP IGR~J17511-3057, observed by \xmm{} for 70 ks and \rxte{} \cite{Papitto2010}, shows both a broad iron line and the Compton hump at $\sim 30$ keV. In this case, the inner disc radius was at $\ge 40$ km from the NS center, assuming a $1.4\, M_\odot$ NS, with an inclination angle between $38^\circ$ and $68^\circ$ (see also \cite{Papitto2016}). 
The AMXP and transitional pulsar IGR~J18245-2452 observed by \xmm{} \cite{Papitto2013b}, showed a broad iron line at 6.7 keV (identified as K$\alpha$ emission from Fe XXV) with a width of $\sim 1.6$ keV, corresponding to R$_{in} \simeq 17.5$ R$_g$ (where $R_g = G M_{NS}/c^2$ is the gravitational radius) or $\sim 36.7$ km for a $1.4\, M_\odot$ NS. For comparison, the inner disk radius derived from the blackbody component was quite similar , $28 \pm 5$ km. The (intermittent) AMXP HETE J1900-2455, observed by \xmm{} for $\sim 65$ ks \cite{Papitto2013a}, showed a broad iron line at 6.6 keV (Fe XXIII-XXV) and an intense and broad line at $\sim 0.98$ keV, visible both in the pn and in the RGS spectrum, compatible with being produced in the same disk region (see Fig.~\ref{fig:spec}, right panel). In this case, the inner disc radius was estimated to be $25 \pm 15\, R_g$, with an inclination angle ranging between $27^\circ$ and $34^\circ$.
A moderately broad, neutral Fe emission line has been observed during the 2015 outburst of IGR~J00291+5934 observed by \xmm{} and \nustar{} \cite{Sanna2017d}. Fitted with a Gaussian profile the line centroid was at an energy of $6.37 \pm 0.04$ keV with a $\sigma = 80 \pm 70$ eV, while using a {\it diskline} profile, the line parameters were poorly constrained. The newly discovered AMXP, MAXI J0911-655, observed by \xmm{} and \nustar{} \cite{Sanna2017d}, shows the presence of a weak, marginally significant and relatively narrow emission line in the range $6.5-6.6$ keV, modelled with a Gaussian profile with $\sigma$ ranging between 0.02 and 0.2 keV, identified with a K$\alpha$ transition from moderate-to-highly ionized iron.

The (intermittent) AMXP SAX J1748.9-2021, observed by \xmm{} for $\sim 115$ ks and \inte{} \cite{Pintore2016}, was caught at a relatively high luminosity of $\sim 5 \times 10^{37}$ erg/s corresponding to $\sim 25\%$ of the Eddington limit for a $1.4\, M_\odot$ NS, and, exceptionally for an AMXP, showed a spectrum compatible with a soft state. The broad-band spectrum is in fact dominated by a cold thermal Comptonization component (electron temperature of $\sim 2$ keV) with an additional hard X-ray emission described by a power-law (photon index $\Gamma \sim 2.3$), typically detected in LMXBs in the soft state (see e.g. \cite{DiSalvo2000}). In addition, a number of broad (Gaussian $\sigma = 0.1 - 0.4$ keV) emission features, likely associated to reflection processes, have been observed in the \xmm{} spectrum. A broad iron line was observed at an energy of $\sim 6.7-6.8$ keV, consistent with a Fe XXV K$\alpha$ transition produced in the disc at a distance of $\sim 20-43\, R_g$ ($\sim 42 - 90$ km), with an inclination angle of $\sim 38-45^\circ$. The other broad emission lines may be associated to K-shell emission of S XVI (2.62 keV), Ar XVIII (3.32 keV) and Ca XX or Ca XIX (4.11 keV or 3.90 keV, respectively), and are compatible with coming from the same emission region as the iron line. 

High-quality X-ray spectra of \saxj{} were obtained during the 2008 outburst with \xmm{} \cite{Papitto2009} and \suzaku{} \cite{Cackett2009} and during the 2015 outburst with \xmm{} (which observed the source at the peak of the outburst) and \nustar{} which observed the source four days later \cite{DiSalvo2019}. The 2015 spectrum of \saxj{} taken with \nustar{} was quite similar to the 2008 spectra; the continuum emission was modelled with one or two blackbody-like components and a hard Comptonization component with an electron temperature $> 40$ keV. On the other hand, the 2015 \xmm{} spectrum was surprisingly much softer, with an electron temperature below 10 keV and a much colder blackbody component (corresponding to a large radius, $> 100$ km, for the emitting surface, \cite{DiSalvo2019}). 
In all the cases, a reflection component was also required to model both the broad iron line and the Compton hump observed on top of the continuum (the composite broad-band spectrum of \saxj{} observed during the 2015 outburst is shown in Fig.~\ref{fig:spec}, left panel). All the smearing parameters were quite similar in the 2008 and 2015 spectra, with the exception of the ionization parameter, much higher during the 2015 \xmm{} observation ($\log \xi \sim 3.5$),
which also showed broad emission lines from highly ionized elements (S XVI, Ar XVIII and Ca XIX-XX) at low energies. In particular, the inner disk radius was $\sim 10\, R_g$, corresponding to about 20 km; for comparison the co-rotation radius of the system is $31 m_{1.4}^{1/3}$ km, where $m_{1.4}$ is the NS mass in units of $1.4\, M_\odot$, indicating that the disk is truncated inside the co-rotation radius during the outburst, as expected from the observed timing properties of the source. 
However, the inclination angle is required to be high (usually values $> 60^\circ$ are required; the best constraint comes from the 2015 \xmm{} spectrum, where $i = 58^\circ-64^\circ$, \cite{DiSalvo2019}). This result is in agreement with evidences from the 2015 \xmm{} spectrum of discrete absorption features, namely an absorption edge at $\sim 7.4$ keV from neutral or mildly ionized iron and at least two absorption lines, possibly from K transitions of highly ionized (He-like) Ne IX (at 0.947 keV) and Mg XI (at 1.372 keV). These lines are relatively broad (implying a velocity dispersion of $\sigma_v \sim 0.01c$) and blue-shifted at a velocity a few percent the speed of light \cite{DiSalvo2019}. If confirmed, these lines may suggest the presence of a weakly relativistic outflowing wind towards the observer. A high inclination angle is also compatible with other estimates (see e.g. \cite{Ibragimov2009,Deloye2008,Bildsten2001,Wang2013}). However, high values for the inclination angle of the system look at odd when considered together with optical estimates of the radial velocity of the companion star \cite{Elebert2008,Cornelisse2009}, since it implies quite low values for the NS mass, $M_{NS} \sim 0.5-0.8\, M_\odot$. These results may indicate some problem with the interpretation of the reflection component and/or the need of more precise measurements of the radial velocity of the companion star. 

It is noteworthy that reflection features are not always observed in AMXPs. Indeed, no evidence of iron emission lines or reflection humps has been reported for IGR~J16597-3704 \cite{Sanna2018a} observed by \swift{} and \nustar{}, IGR~J17379-3747 \cite{Sanna2018b} observed by \xmm{}, XTE J1807-294 \cite{Falanga2005a} observed by \rxte{}, \xmm{} and \inte{}, and XTE J1751-305 \cite{Miller2003} observed by \xmm{}. Similar results have been reported for the 2018 outburst of the AMXP SWIFT J1756.9-2508 monitored by several satellites such as \nicer{}, \swift{}, \xmm{}, \nustar{} and \inte{}.  Evidences of iron emission lines in the 6 -- 7 keV band have been, however, reported from the analysis of \rxte{} observations of the source during its 2007 and 2009 outbursts \cite{Patruno2010c}.

\subsection{Short-term variations of the spin during outbursts}

Accretion torque theories can be tested studying the spin variations of AMXPs during accretion phases. These studies can provide valuable information on the mass accretion rate and magnetic field of the NS in these systems, as well as their spin evolution. Coherent timing has been performed on several sources of the sample, with sometimes controversial results (see \cite{Campana2018} for a recent review). Some sources seem to show spin-up during outbursts (e.g.\ IGR~J00291+5934, XTE J1751-305, XTE J1807-294, IGR~J17511-3057), while other sources seem to show spin-down even during accretion phases (e.g. XTE J0929-314,  XTE J1814-338, IGR~J17498-2921 and IGR~J17591-2342). Although some AMXPs show pulse phase delays distributed along a second order polynomial, indicating an almost constant spin frequency derivative, other sources show strong timing noise (e.g. SAX J1808-3658, HETE 1900-2455), sometimes correlated with sharp variations of the X-ray flux, which can hamper any clear measurement of the spin derivative.

\begin{figure}
\begin{subfigure}{.5\textwidth}
  \centering
  \includegraphics[width=1.0\linewidth]{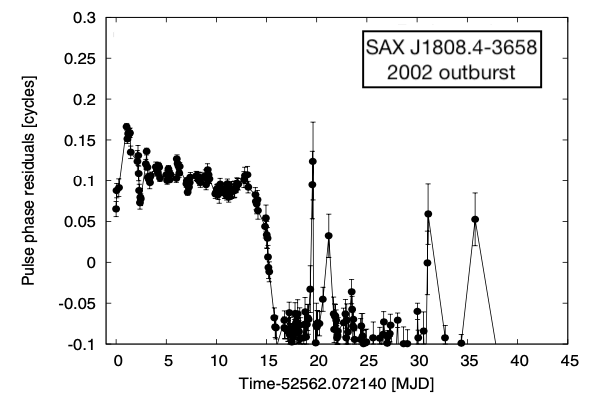}
  \label{fig:sfig1}
\end{subfigure}%
\begin{subfigure}{.5\textwidth}
  \centering
  \includegraphics[width=1.05\linewidth]{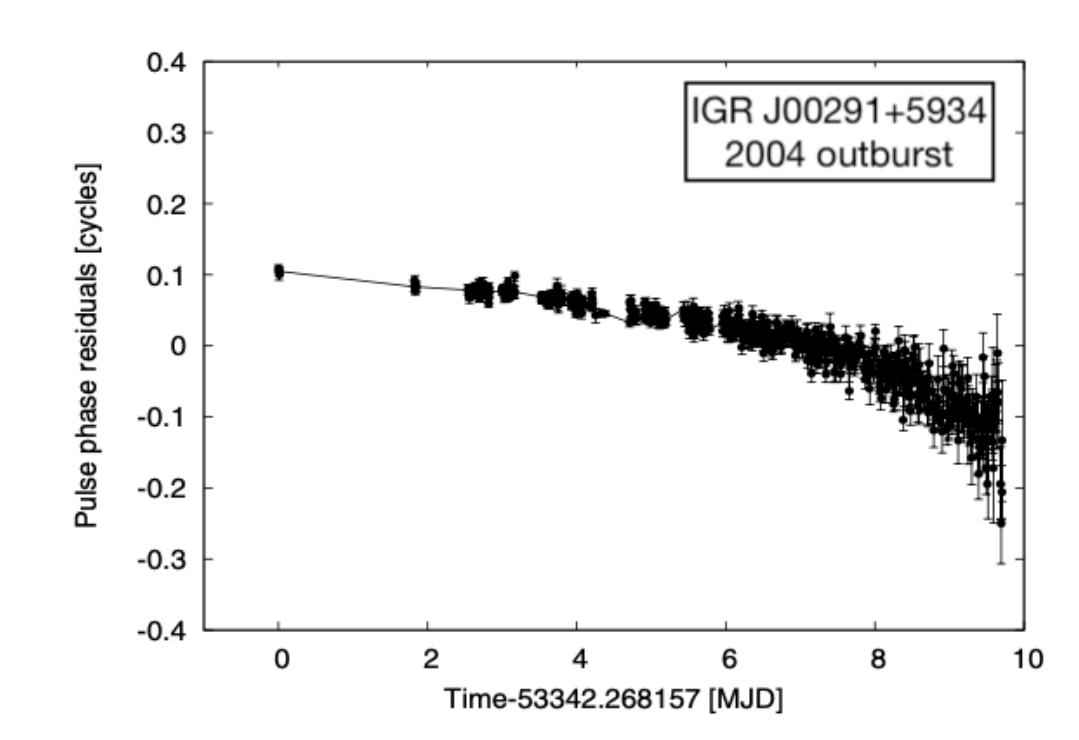}
  \label{fig:sfig2}
\end{subfigure}
\caption{Phase residuals of the fundamental frequency for the AMXP \saxj{} (left Panel) and IGR~J00291+5934 (right Panel) [Figure adapted from \cite{Patruno2009d}]}
\label{fig:tim_noise}
\end{figure}

The first AMXP for which a spin derivative has been measured is the fastest spinning ($\sim 599$ Hz, in a 2.46 hr orbit) among these sources, IGR~J00291+5934. It is now generally accepted that this source shows spin up at a rate of $\sim (5-8) \times 10^{-13}$ Hz s$^{-1}$ \cite{Falanga2005b,Patruno2010,Hartman2011,Papitto2011c} (see Fig.~\ref{fig:tim_noise}, right panel). \cite{Burderi2007} have attempted to fit the phase delays vs.\ time with physical models taking into account the observed decrease of the X-ray flux as a function of time during the X-ray outburst, with the aim to get a reliable estimate of the mass accretion rate onto the compact object. 
Because the X-ray flux, which is assumed to be a good tracer of the mass accretion rate, is observed to decrease along the outburst, this variation has to be considered in eq.(\ref{torque}) in order to obtain the correct value of the mass accretion rate, and hence of the spin frequency derivative, at the beginning of the outburst as well as its variation during the outburst. This approach has been successfully applied to the timing of the 2014 outburst of the so-called {\it bursting pulsar}, GRO J1744-28, an X-ray pulsar with a spin frequency of $2.14$ Hz in a $11.83$ days orbit around a $0.2\, M_\odot$ companion star. \cite{Sanna2017b} were able in this way to obtain a good fit of the pulse phase delays versus time, deriving a frequency spin-up of $\sim 9 \times 10^{-12}$ Hz/s and inferring a distance to the source between 3.4 and 4.1 kpc, assuming a disk viscous parameter $\alpha$ in the range of 0.1-1.   
In the case of IGR~J00291+5934, this technique gives a spin frequency derivative at the beginning of the outburst of $\dot \nu \sim 1.2(2) \times 10^{-12}$ Hz s$^{-1}$, corresponding to a mass accretion rate of $\sim 7 \times 10^{-9}\, M_\odot/yr$ and a peak bolometric luminosity of $\sim 7 \times 10^{37}$ erg/s for the 2004 outburst. This is at least one order of magnitude higher than the X-ray luminosity inferred from the observed X-ray flux, assuming a distance of 4.2 kpc. Once we will have a direct, independent, estimate of the distance to the source, we will have the possibility to test the $\dot M$ vs.\ X-ray luminosity relation, torque theories and/or the physical parameters of the NS in this system. 

A recent example of an AMXP showing spin-down while accreting is given by IGR~J17591-2342. This source has been extensively monitored by \nicer{} during its outburst starting from 2018 August 15 up to 2018 October 17 for a total exposure time of  $\sim101$ ks distributed into 37 dedicated observations. X-ray pulsations have been detected uninterruptedly for almost 60 days allowing to accurately investigate the NS spin frequency evolution. Phase-coherent timing analysis of the frequency fundamental and second harmonic components revealed a statistically significant negative frequency derivative $\dot{\nu}\sim -7\times 10^{-14}$ Hz/s \cite{Sanna2020c} (see however, \cite{Kuiper2020} for different results from the timing analysis). Further analysis of the the X-ray pulse phase evolution of IGR~J17591-2342, adopting a physical model that accounts for the accretion material torque as well as the magnetic threading of the accretion disc in regions where the Keplerian velocity is slower than the magnetosphere velocity, allows to constrain the NS magnetic field to be $B_{eq} = 2.8(3)\times10^8$ G \cite{Sanna2020c}.

A similar spin frequency evolution has been reported for the AMXPs IGR~J17498-2921 ($\dot{\nu}=-6.3(1.9)\times 10^{-14}$ Hz/s \cite{Papitto2011b}), XTE J1814-338 ($\dot{\nu}=-6.7(7)\times 10^{-14}$ Hz/s \cite{Papitto2007}), and XTE J0929-314 ($\dot{\nu}=-9.2(4)\times 10^{-14}$ Hz/s \cite{Galloway2002}). 
These observations indicate that spin-down during accretion phases is possible and requires a magnetic threading of the accretion disk.  

The best studied, as well as most discussed source, is certainly the first discovered AMXP, SAX J1808-3658. Differently from all the other AMXPs, this source shows X-ray outbursts almost regularly every 2 -- 4 years. To date we have observed (with instruments with high-time resolution capabilities) eight outbursts from this source, each lasting about one month. The pulse phase evolution during the outburst shows a strong timing noise, with phases going up and down without any clear correlation with flux variations, or remaining constant for a long time before jumping to another constant phase value (see e.g. \cite{Hartman2008,Hartman2009a, Hartman2009b}). In the attempt to gain information on the spin variations in this source, \cite{Burderi2006} have analysed separately the fundamental and second harmonic of the pulse profile during the 2002 outburst of the source, finding that while the phases of the fundamental are dominated by timing noise (see Fig.~\ref{fig:tim_noise}, left panel), the second harmonic shows a more regular behaviour. This suggests that the phase jump in the fundamental (clearly visible in Fig.~\ref{fig:tim_noise}, left panel) is not related to an intrinsic spin variation (which would have affected the whole pulse profile), but is instead caused by a change of the shape of the pulse profile, leaving the possibility that the harmonic is a better tracer of the NS spin. A similar behavior of the second harmonic has also been observed in other AMXPs (e.g. \cite{Riggio2008,Riggio2011,Papitto2012}).  Under this hypothesis, the fitting of the second harmonic phase delays reveals a spin-up at the beginning of the outburst of $\dot \nu = 4.4(8) \times 10^{-13}$ Hz s$^{-1}$, corresponding to a mass accretion rate of $\dot M \sim 1.8 \times 10^{-9}\, M_\odot$ yr$^{-1}$, and a constant spin-down, of $\dot \nu_{sp} = -7.6(1.5) \times 10^{-14}$ Hz s$^{-1}$, dominating the phase delays at the end of the outburst. The mass accretion rate inferred from timing is only a factor of 2 larger than the inferred X-ray bolometric luminosity at the beginning of the outburst, that is $\sim 10^{37}$ ergs s$^{-1}$. The spin-down can be interpreted in terms of the threading of the accretion disk by the NS magnetic field outside the co-rotation radius, which, in agreement with expectations, appears to be more relevant at the end of the outburst, when the mass accretion rate significantly decreases. The derived magnetic field, $B = (3.5 \pm 0.5) \times 10^8$ G, is perfectly in agreement with other, independent, constraints (see \cite{Burderi2006}, and references therein).

The latest outburst of \saxj{} in 2019 was monitored with \nicer{} for one month and a total exposure of 355.4 ks. Timing analysis of these data showed that the pulse profile was dominated by the fundamental (the second harmonic was significantly detected only in a handful of intervals) and the relative phase delays show a clear parabolic trend typical of a spin-down at the rate of $\dot{\nu} = -3.02(13)\times 10^{-13}$ Hz s$^{-1}$ \cite{Bult2020}. Since these phase shifts appear to be correlated with the source flux, the authors interpreted this trend in terms of hot-spot drifts on the stellar surface, driven by changes in the mass accretion rate.    

Other recent results regard phase-coherent timing analysis of the outburst of the AMXPs SWIFT~J1759-2508 and IGR~J17379-3737, which allowed to set upper limits on the spin frequency derivative of $\dot{\nu}<|1.4|\times 10^{-12}$ Hz/s \cite{Sanna2019,Bult2018b} and $-0.5\times 10^{-14}<\dot{\nu}<0.9\times 10^{-14}$ Hz/s \cite{Bult2020}, respectively.

\subsection{Long-term variations of the spin}
AMXPs for which more than one outburst has been observed with high time resolution instruments, allow to derive long term spin evolution comparing the averaged spin frequency measured in each outburst. To date only six AMXPs have been observed in different outbursts: SAX J1808.4-3658, IGR~J00291+5934, XTE J1751-305, Swift J1756.9-2508, IGR~J17379-3747, IGR~J17511-3057, NGC 6440 X-2, and SAX J1748.9-2021 (although, with relatively low S/N and short outburst duration for some of these sources, see Table ~\ref{Tab2}). 

\begin{figure}
\begin{subfigure}{.5\textwidth}
  \centering
  \includegraphics[width=1.0\linewidth]{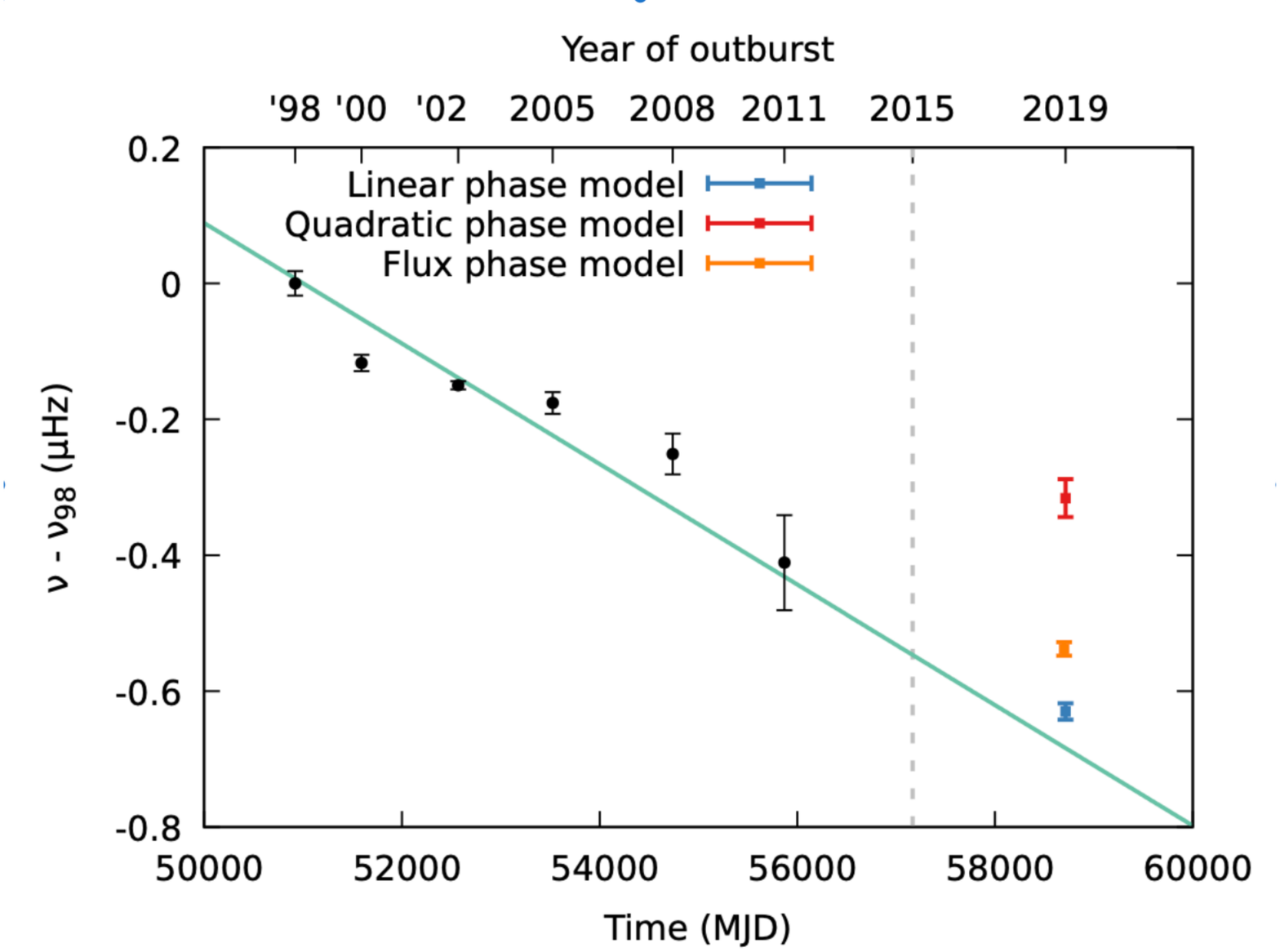}
  \label{fig:sfig1}
\end{subfigure}%
\begin{subfigure}{.5\textwidth}
  \centering
  \includegraphics[angle=-90,width=1.3\linewidth]{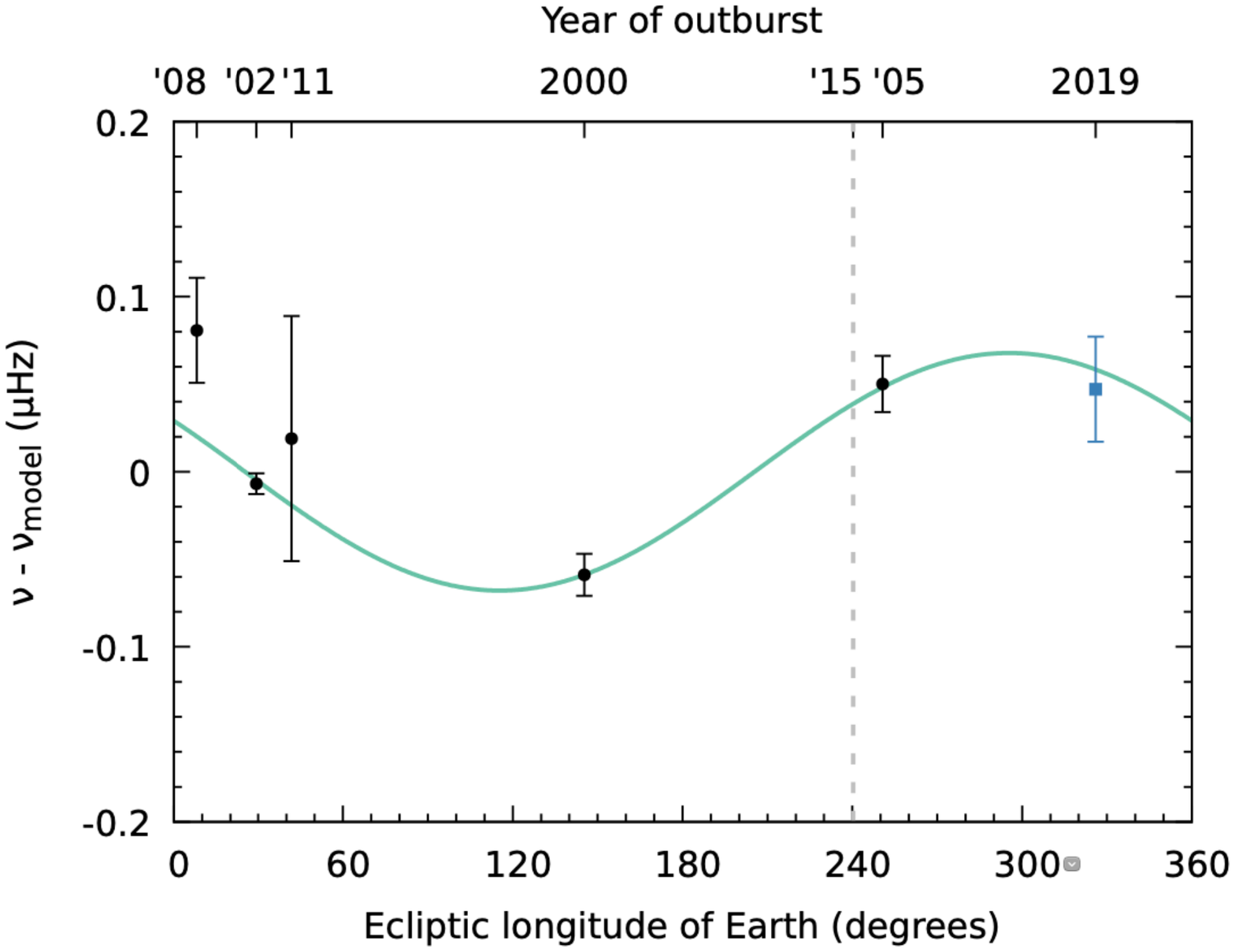}
  \label{fig:sfig2}
\end{subfigure}
\caption{\textit{Left Panel:} Secular spin frequency evolution of the AMXP \saxj{} calculated relatively to the 1998 epoch. Black points represent measurements obtained with \rxte{} while colored squared represent the \nicer{} measurements obtained for the 2019 outburst of the source for three different models (see \cite{Bult2020} for more details). The solid line indicates the spin evolution best-fit model. \textit{Right Panel:} Spin frequency measurements of the AMXP \saxj{} relative to the best-fit spin-down model as a function of the Earth's ecliptic longitude. [Figure from \cite{Bult2020}]}
\label{fig:1808_sec}
\end{figure}

The best constrained long-term spin evolution is obtained for \saxj{} (see left panel of Fig.~\ref{fig:1808_sec}), for which secular spin evolution has been measured over a 13 year time span (between 1998 and 2011), which shows a constant long-term spin-down at a rate of $\sim -1 \times 10^{-15}$ Hz s$^{-1}$ (see \cite{Patruno2012}, and references therein). Because of the stability of the spin-down rate over the years, the most likely explanation appears to be loss of angular momentum via magnetic-dipole radiation, which is expected for a rapidly rotating NS with a magnetic field. The measured spin-down is consistent with a polar magnetic field of $(1.5 - 2.5) \times 10^8$ G, in agreement with other estimates. The spin frequency measured during the 2015 outburst had a large uncertainty because of strong timing noise of the fundamental. Interestingly, the spin frequency measured using the phases of the second harmonic falls very close (less than $2 \sigma$) from the value predicted by the secular evolution (see \cite{Sanna2017c}). For the 2019 outburst, the second harmonic is significantly detected only in few \nicer{} snapshots and the exact value of the spin frequency inferred from the fundamental depends on the adopted timing solution. \cite{Bult2020} have fitted the phase delays using a linear model (which leaves large residuals), a quadratic model (indicating a spin-down during the outburst), and a flux-adjusted model (under the hypothesis that phase variations with time originate from a hot-spot drifting on the stellar surface, driven by changes in the mass accretion rate). The linear and flux-adjusted models give a spin frequency relatively close to the secular spin-down trend, while the quadratic model gives a frequency lying significantly above the trend (see Fig.~\ref{fig:1808_sec}, left panel). Considering the linear model (which provides the frequency value closest to the expected trend), the long-term evolution of the spin shows a modulation around a constant spin-down behaviour at the Earth's orbital period (right panel of Fig.~\ref{fig:1808_sec}), which is used to astrometrically refine the source coordinates.

A long-term spin-down has also been measured for IGR~J00291+5934 between the 2004 and 2008 outbursts, at a rate of $-4.1(1.1) \times 10^{-15}$ Hz s$^{-1}$ \cite{Papitto2011c,Patruno2010,Hartman2011}, larger than that observed in \saxj{}, as expected given that IGR~J00291+5934 spins at a higher frequency. 
The less accurate spin measurement from the \xmm{} observation of
its 2015 outburst could only constrain the spin-down since the
previous outburst as $|{\nu}|<6\times 10^{-15}$~Hz~s$^{-1}$ (see
[113] and references therein.
If interpreted in terms of magneto-dipole emission, the measured spin-down translates into an estimate of the NS magnetic field of $(1.5-2) \times 10^8$ G. Another possibility is given by the spin-down torque associated with the emission of GR, strongly dependent on the NS spin, which has also been proposed as a limiting factor for the maximum spin frequency observed for a NS (to date 716 Hz, \cite{Hessels2006}). Assuming that the long-term spin-down observed in IGR~J00291+5934, the fastest spinning AMXP known to date, is due to this mechanism, the measurement of the average spin-down in this source translates to an upper limit on the average mass quadrupole moment of $Q \lesssim 2 \times 10^{36}$ g cm$^2$ \cite{Hartman2011}. Under this hypothesis, it is possible to predict that the long-term spin-down in IGR~J00291+5934 should be a factor 7.6 higher than in \saxj{}. The large uncertainties on these measurements prevent at the moment to assess this prediction, but it can be checked with future, high-quality, monitoring of the spin frequency in these systems.

Long-term spin evolution has been constrained for a few other sources of the sample of AMXPs. After the discovery of X-ray pulsations during the 2018 outburst of IGR~J17379-374, pulsations from this source have been discovered also in the \rxte{} archival data of its 2004 and 2008 outbursts after applying the binary ephemeris. Combining the barycentric spin frequency values from the three oubursts, an upper limit on the secular spin derivative has been estimated, $-8.3\times10^{-13}$ Hz/s $<\dot{\nu}<1.1\times 10^{-12}$ Hz/s. This corresponds to an upper limit on the magnetic field strength of $B<2.8\times 10^9$ G, under the assumption of a NS radius of 10 km and an angle $\alpha\simeq 10^\circ$ between the magnetic hotspot and the rotational pole \cite{Sanna2018b}. Swift J1756.9-2508 has been detected in outburst three times (2007, 2009 and 2018) since its discovery, which allowed the detection of a long-term spin-down derivative of $-4.8(6)\times 10^{-16}$ Hz/s \cite{Sanna2019}, corresponding to a NS superficial magnetic field $1.5\times 10^8 < B_{eq} < 2.2\times 10^8$ G (consistent with the value reported by \cite{Mukherjee2015}).

\subsection{Long-term timing of the orbital period}
The study of the orbital evolution in Low Mass X-ray Binary systems is very important to constrain the evolutionary path leading to the formation of rotation-powered MSPs, and hence to obtain information on the progenitors of fast-rotating NS and on the recycling scenario. It is worth noting, however, that the discussions on the AMXPs long-term changes of the orbital period described in this section reflect changes on timescales relatively short with respect to the secular evolution of the binary systems (see Chapter~9 for more details on the topic). Nonetheless, orbital evolution can in principle be useful to put constraints on alternative theories of Gravity. In fact, since the difference in the orbital period evolution of binaries interpreted with General Relativity (GR) and other theories of Gravity (e.g. Brans-Dicke gravity) is related to the mass difference of the two members of the binary system \cite{Will2006}, these sources provide prime candidates for constraining deviations from GR \cite{Psaltis2008}. In this framework, AMXPs are the most promising candidates for an experimental test on these alternative theories, because the companion star is, in most cases, a very light white dwarf or even a brown dwarf \cite{Bildsten2001}, and the primary stars are millisecond pulsars with orbital periods accurately determined. 
 
However, these studies require a large baseline (tens of years) of data to be able to constraint the orbital period derivative. Hence, one of the main difficulty is given by the limited number of AMXPs observed recurrently into X-ray outburst. To date, only eight AMXPs have more than one outburst observed with high-time resolution instruments since their discovery, and therefore only few constraints on the orbital period derivative have been derived to date (see Table ~\ref{Tab2}). Moreover, long-term orbital solutions show sometimes residuals that are complex and difficult to interpret. {Understanding these complex orbital residuals is therefore of fundamental importance, since it would allow to constrain the orbital period evolution in these systems, at least on a dynamical timescale, 
providing hints on their evolutionary paths or at least important information on the long-term dynamical behaviour of these systems.
Furthermore, the precise determination of the orbital period derivative caused by mass transfer may give in perspective the possibility to constrain alternative theories of Gravity.

The best constraint on the orbital evolution of AMXPs comes again from \saxj{}, which has shown eight X-ray outbursts to date, allowing to follow its orbital period over 21 years. As reported in the left panel of Fig.~\ref{fig:orb_evo}, for each outburst the time of passage of the NS through the ascending node ($T^*$, which is the most sensitive parameter to variations of the orbital period) can be derived and plotted versus time. The orbital residuals (with respect to a constant orbital period) were dominated by a clear parabolic trend up to the 2015 outburst, with residuals with respect to this trend of the order of few seconds \cite{Sanna2017c}. Interpreting this parabolic trend as a constant orbital period derivative, the best-fit value is $\dot P_{orb} = 3.6(4) \times 10^{-12}$ s s$^{-1}$, implying a strong orbital expansion. The origin of the observed $\dot P_{orb}$ is still not fully understood, yet, and different possible mechanisms have been proposed over the years (see e.g. \cite{DiSalvo2008,Hartman2008,Burderi2009,Patruno2012b,Patruno2016}. However, there is consensus on the fact that conservative mass transfer is not compatible with the observed value of $\dot P_{orb}$ for \saxj{}. This can be easily demonstrated by estimating the mass-loss rate from the secondary star as a function of the observed orbital period derivative (see e.g. \cite{Burderi2009}), which implies a mass transfer of the order of $2 \times 10^{-9}\, M_\odot$ yr$^{-1}$. This mass transfer rate is much larger than the mass accretion rate onto the NS, considering that the source accretes matter for about a month every 2-4 yr with a bolometric luminosity at the peak of the outburst barely reaching $10^{37}$ erg s$^{-1}$ (corresponding to a maximum mass accretion rate of $\sim 10^{-9}\, M_\odot$ yr$^{-1}$). The average mass accretion rate over the 17 years from 1998 to 2015 is indeed three orders of magnitude below, $\sim 2 \times 10^{-11}\, M_\odot$ yr$^{-1}$. 

A not conservative mass transfer can explain the large orbital period derivative assuming that the mass transfer rate is $\dot M \sim 10^{-9}\, M_\odot$ yr$^{-1}$, and that most of the transferred matter is expelled from the system, instead of being accreted onto the NS, with the specific angular momentum at the inner Lagrangian point (see \cite{DiSalvo2008,Burderi2009}). In this case, the non-conservative mass transfer may be a consequence of the so-called \textit{radio-ejection} model, extensively discussed by \cite{Burderi2001}, envisaging that a fraction of the transferred matter in the disc could be swept out by the (radiative and high-energy particles) pressure of the pulsar wind. Alternatively, the large orbital period derivative observed in \saxj{} has been interpreted as the effect of short-term angular momentum exchange between the donor star and the orbit \cite{Hartman2009b,Patruno2012b}, resulting from variations in the spin of the companion star (holding the star out of synchronous rotation) caused by intense magnetic activity driven by the pulsar irradiation, the so-called Applegate \& Shaham mechanism (hereafter A\&S \cite{Applegate1994}). In this case, the orbital period should oscillate, alternating epochs of orbital period increase and decrease, because of the gravitational quadrupole-coupling to the orbit. However, according to this mechanism, the system should evolve to longer orbital periods, because of mass and angular momentum loss, on a timescale of $10^8$ yr (for a 2-hr orbital period and a companion mass of $0.1-0.2\, M_\odot$), thus implying a strong orbital period derivative, similar to that inferred from the quadratic trend observed in \saxj{}. In this framework, the orbital residuals in \saxj{} up to 2015 may be interpreted as small oscillations of few-seconds amplitude caused by the A\&S mechanism superposed on a global orbital period derivative induced by the strong mass-loss from the system \cite{Sanna2017c}. Alternatively, variations of the orbital period with respect to the global parabolic trend may be caused by random fluctuations of the mass transfer (and loss) rate. 

The latest outburst of \saxj{} in 2019, however, shows an abrupt flattening of the parabolic trend \cite{Bult2020} (as is evident in Fig.~\ref{fig:orb_evo}, left, top panel). Indeed, the measurements between 2008 and 2019 taken alone seem to imply an orbital contraction of the orbit in the last 10 years, with an orbital period derivative of $\dot P_{orb} \simeq -5.2 \times 10^{-12}$ s s$^{-1}$. Alternatively, fitting all the measurements with a global parabolic trend, gives an orbital period derivative of $\dot P_{orb} = 1.6 \pm 0.7 \times 10^{-12}$ s s$^{-1}$. As shown in the left, bottom panel of Fig.~\ref{fig:orb_evo}, the residuals around this mean trend show a sinusoidal-like, 7-s amplitude, oscillation with a period of approximately 18.2 years. Additional monitoring of future outbursts is needed to confirm the presence of oscillations around a steadily expanding orbit, or, instead, a $\sim 20$ s amplitude modulation around a constant (or much less variable) orbital period.

\begin{figure}
\begin{subfigure}{.5\textwidth}
  \centering
  \includegraphics[width=1.0\linewidth]{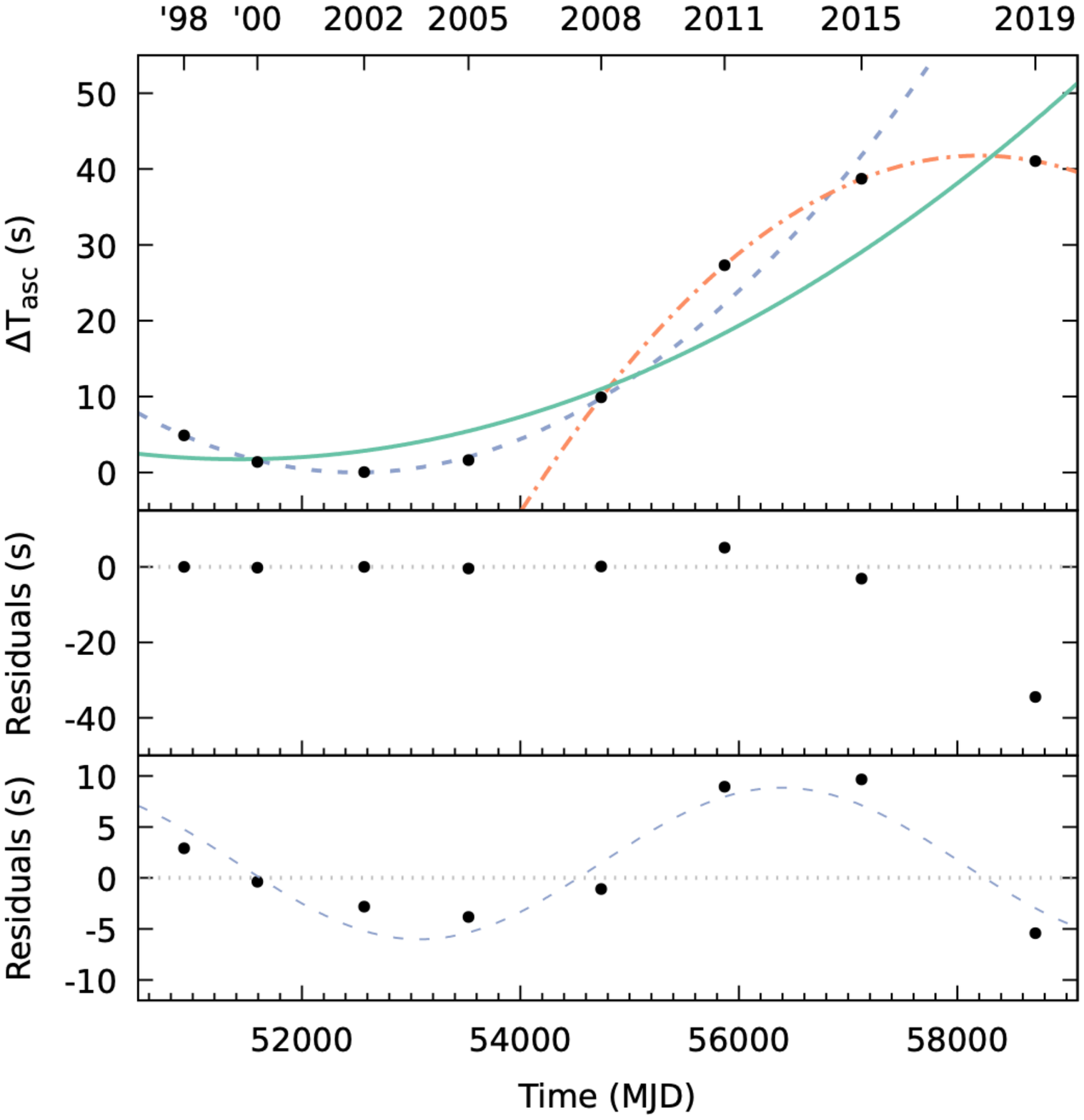}
  \label{fig:sfig1}
\end{subfigure}%
\begin{subfigure}{.5\textwidth}
  \centering
  \includegraphics[width=1.45\linewidth]{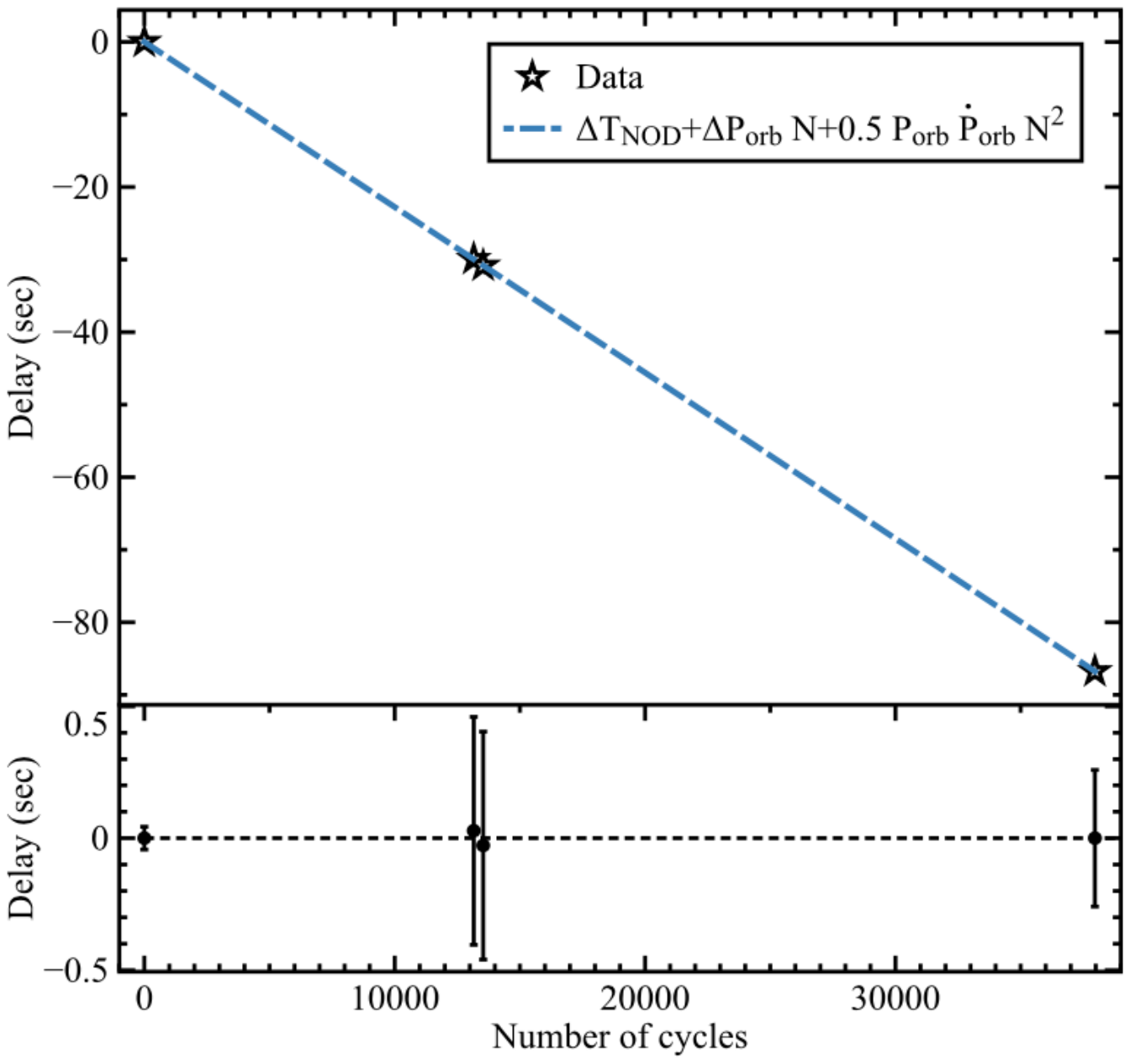}
  \label{fig:sfig2}
\end{subfigure}
\caption{\textit{Left Panel:} Orbital evolution of the AMXP \saxj{}. The dashed, dashed-dot and solid curves represent the parabolic trends fit between 1998-2008, 2008-2019, and 1998-2019 subsets of the data, respectively. Residuals relative to the 1998-2008 and 1998-2019 parabolic models are shown in the middle and bottom panels, respectively. The dashed line shown in the bottom panel represents a sinusoid with a 18.2 yr period and 7 s amplitude that has been inserted to tentatively describe the residuals (see \cite{Bult2020} for more details). The solid line indicates the spin evolution best-fit model. \textit{Right Panel:} Orbital evolution of the AMXP IGR~J00291+5934. The cyan dashed line represents the best-fitting parabola used to model the data. Residuals in seconds of the time delays with respect to the best-fitting timing solution are shown in the bottom panel. [Figures from \cite{Bult2020,Sanna2017d}]}
\label{fig:orb_evo}
\end{figure}

A very different evolution is found for IGR~J0029+5934, as shown in the right panel of Fig.~\ref{fig:orb_evo}, which has orbital parameters very similar to those of \saxj{}, and it is considered its orbital twin. IGR~J0029+5934 has shown only four outburst since its discovery, but tight upper limits could be derived on its orbital period derivative, $|\dot P_{orb}| < 5 \times 10^{-13}$ s s$^{-1}$ (90\% confidence level \cite{Patruno2017,Sanna2017d}). This implies a much slower orbital evolution, on a timescale longer than $\sim 0.5$ Gyr, as compared to the fast (up to 2015) orbital evolution of \saxj{}, $\sim 70$ Myr. Although the orbital evolution observed in IGR~J0029+5934 is obtained using only four points with large error bars, and more measurements are needed to confirm this result, it seems to be compatible with the expected timescale of mass transfer driven by angular momentum loss via GR, with no need of A\&S mechanism or non-conservative mass transfer. 

\begin{table}
\caption{Accreting Millisecond X-ray Pulsars: secular spin and orbital evolution}
\scriptsize
\begin{center}
\begin{tabular}{lcccccl}
\hline
\hline
Source & \# outbursts & $P_{\rm orb}$ & $T_{ASC}$  & $\dot{P}_{\rm orb}$ & $\dot{\nu}$ & Ref.\\
 &  & (s) & (MJD) & (s/s) & (Hz/s) &\\
\hline
\textbf{AMXP} & & & & & & \\
\hline
NGC 6440 X-2 & 4 & 3457.8929(7) & 55318.04809(2) &  $\pm8\times 10^{-11}$ &$\pm5\times 10^{-13}$ & \cite{Bult2015c}\\
SAX J1748.9-2021 & 4 & 31554.868(3) & 52191.52737(3) &  $3.9(1.1)\times 10^{-11}$ & -& \cite{Sanna2020}\\
IGR J00291+5934  & 4 & 8844.07673(3) &53345.16192640(5)  & $-0.7(2.2)\times 10^{-13}$ & $-3.0(8)\times 10^{-15} $  &  \cite{Patruno2017,Sanna2017d,Patruno2010,Papitto2011c}\\
IGR J17379-3747  & 3 & 6765.84521(3) & 53056.03926(12) &  $−2.5(2.3)\times 10^{-12}$ & -  & \cite{Sanna2018b}\\
SAX J1808.4-3658 & 8 & 7249.1541(2) & 50914.79449(4) & $1.7(0.6)\times 10^{-12}$ & $-1.01(7)\times 10^{-15}$ & \cite{Bult2020,Sanna2020b}\\
Swift J1756.9-2508 &3  & 3282.3519(5) &  54265.28087(10) &  $1.5(2.8)\times 10^{-12}$ & $-4.8(6)\times 10^{-16}$ & \cite{Sanna2018d,Bult2018b}\\
IGR J17511-3057 & 2 & 12487.50(7) & 57107.85882(8) & $4.4(7)\times10^{-11}$  & - &  \cite{Riggio2020} \\
IGR J1751-305 & 2 & 2545.342(2) & 52368.0129023(4) & $\pm1.4\times10^{-11}$  & $-5.5(1.2)\times 10^{-15}$ &  \cite{Riggio2011b}\\
\hline
\hline
\end{tabular}\\
\end{center}
$P_{\rm orb}$ is the orbital period, $T_{ASC}$ is the time of passage from the Ascending Node and the reference of the orbital solution, $\dot{P}_{\rm orb}$ is the orbital period derivative, and $\dot \nu$ is the long-term spin frequency derivative.
\label{Tab2}
\end{table}

What causes such an enormous difference between the orbital evolution of two sources with very similar orbital parameters?  
\cite{Tailo2018} have studied the effects of irradiation of the companion star in order to reproduce the evolution of \saxj{}. They have simulated the binary evolution of its possible progenitor system, starting at an orbital period of $\sim 6.6$ h and taking into account angular momentum losses via MB and GR. They also consider the effects of illumination of the donor by both the X-ray emission during accretion phases and the spin-down luminosity of the pulsar. They show that pulsar irradiation is a necessary ingredient to reach the correct orbital period when the donor mass is reduced to the actual value of $0.04-0.06\, M_\odot$. Also it is shown that irradiation alters the internal structure of the donor, causing the companion star to be not completely convective at the values of mass observed for the system and keeping the MB active along the whole evolution (see also \cite{Chen2017}). Mass transfer proceeds through cycles: the donor reacts to the irradiation expanding and starting a phase of large mass-transfer; consequently, mass loss dominates the period evolution. When the thermal relaxation of the envelope takes over, the star radius shrinks and the system detaches (see also \cite{Benvenuto2017} and references therein). In this framework, \saxj{} and IGR~J0029+5934 may be at different phases of this cycling behavior, with the first in a phase of high mass transfer rate (and a fast orbital evolution) and the latter in an almost detached phase (with a low mass transfer rate and slow orbital evolution). In both cases, a non-conservative mass transfer is implied with matter expelled from the system by the radiation pressure of the pulsar, that should be stronger in the case of IGR~J0029+5934 because of its faster rotation. More details on the orbital evolution of these systems from a theoretical point of view can be found in Chapter 9 of this book. 

In order to test this or other models for the orbital evolution in these systems it is important to continue monitoring the behavior of these and other sources. Other AMXPs have shown more than one outburst for which an orbital solution has been obtained. Long-term evolution of the time of passage from the ascending node of SAX J1748.9-2021 has been clearly observed after combining the orbital solutions of the five observed outbursts to date (in 2001, 2005, 2010, 2015 and 2018). Although marginally significant ($\sim 3.5 \sigma$ confidence level), an orbital period derivative of $\dot{P}_{\rm orb}=3.9(1.1)\times 10^{-11}$ s/s has been determined \cite{Sanna2020}, suggesting again a fast orbital expansion of the system. In the case of IGR~J17379-3747, the combination of the ephemerides obtained for the observed outbursts allows to set an upper limit on the orbital period derivative of $-4.4\times10^{-12} < \dot{P}_{\rm orb} < 9.4\times 10^{-12}$ \cite{Sanna2018b}. Swift J1756.9-2508 has been detected in outburst three times (2007, 2009 and 2018) since its discovery; the orbital timing of the source sets an upper limit on the orbital period derivative of $-4.1 \times 10^{-12} < \dot{P}_{\rm orb} < 7.1 \times 10^{-12}$ \cite{Sanna2019}. \cite{Riggio2020} analysed a \nustar{} observation of the 2015 outburst of IGR~J17511-3057, obtaining a new local orbital solution. Combining that with the the orbital solution of the 2011 outburst \cite{Riggio2011}, they inferred an orbital period derivative of $\dot{P}_b = 4.4(7) \times 10^{-11}$ s s$^{-1}$, suggesting a fast orbital expansion of the binary system similar to that reported for SAX J1748.9-2021. These results are summarised in Table ~\ref{Tab2}.

\subsection{Not conservative mass transfer?}

Despite the reduced statistics, the majority of the results suggests that these sources are undergoing a fast orbital expansion, notwithstanding the low averaged mass accretion rate observed from these sources.

Besides AMXPs, one of the most evident example of non-conservative mass transfer is given by the slowly rotating (spin period of $\sim 0.59$ s, \cite{Jonker2001}) X-ray pulsar and eclipsing LMXB 4U 1822-37, which shows a persistent X-ray luminosity of $\sim 10^{36}$ erg/s and an orbital period of $\sim 5.57$ h, measured from the periodic eclipse of the X-ray source and confirmed through the timing of the X-ray pulsations. The compilation of the eclipse times over the last 40 years shows a fast orbital expansion at a rate of $\dot P_{orb} \sim 1.5 \times 10^{-10}\, M_\odot/yr$ (see e.g. \cite{Chou2016,Mazzola2019}). The delays on the eclipse arrival times with respect to a constant orbital period show a clear parabolic trend, which implies a constant orbital period derivative, more than three orders of magnitude what is expected from conservative mass transfer driven by MB and GR (e.g. \cite{Burderi2010,Iaria2011}). Mechanisms based on the gravitational quadrupole coupling of the companion star with the orbit (see e.g. \cite{Applegate1992,Applegate1994}) have been investigated, however, they resulted not suitable since the ($\sim 0.3\, M_\odot$) companion star lacks enough internal power to produce such a large orbital period variation (e.g. \cite{Mazzola2019}). 
A possible explanation is given by a highly not conservative mass transfer, in which the companion star transfers mass at a high rate. Most of the transferred mass is then expelled from the system by the strong radiation pressure of the central source emitting at the Eddington limit. In fact, it has been proposed that 4U 1822-37 is accreting at the Eddington limit (while just 1\% of the total X-ray luminosity is visible due to the high inclination angle, $80-85^\circ$, \cite{Iaria2011}), while the companion star is transferring at a higher rate (of the order of seven times Eddington, \cite{Burderi2010}), and most of the transferred mass is expelled from the system by the radiation pressure producing strong outflows and winds.

\begin{figure}
\begin{subfigure}{.5\textwidth}
  \centering
  \includegraphics[width=1.2\linewidth]{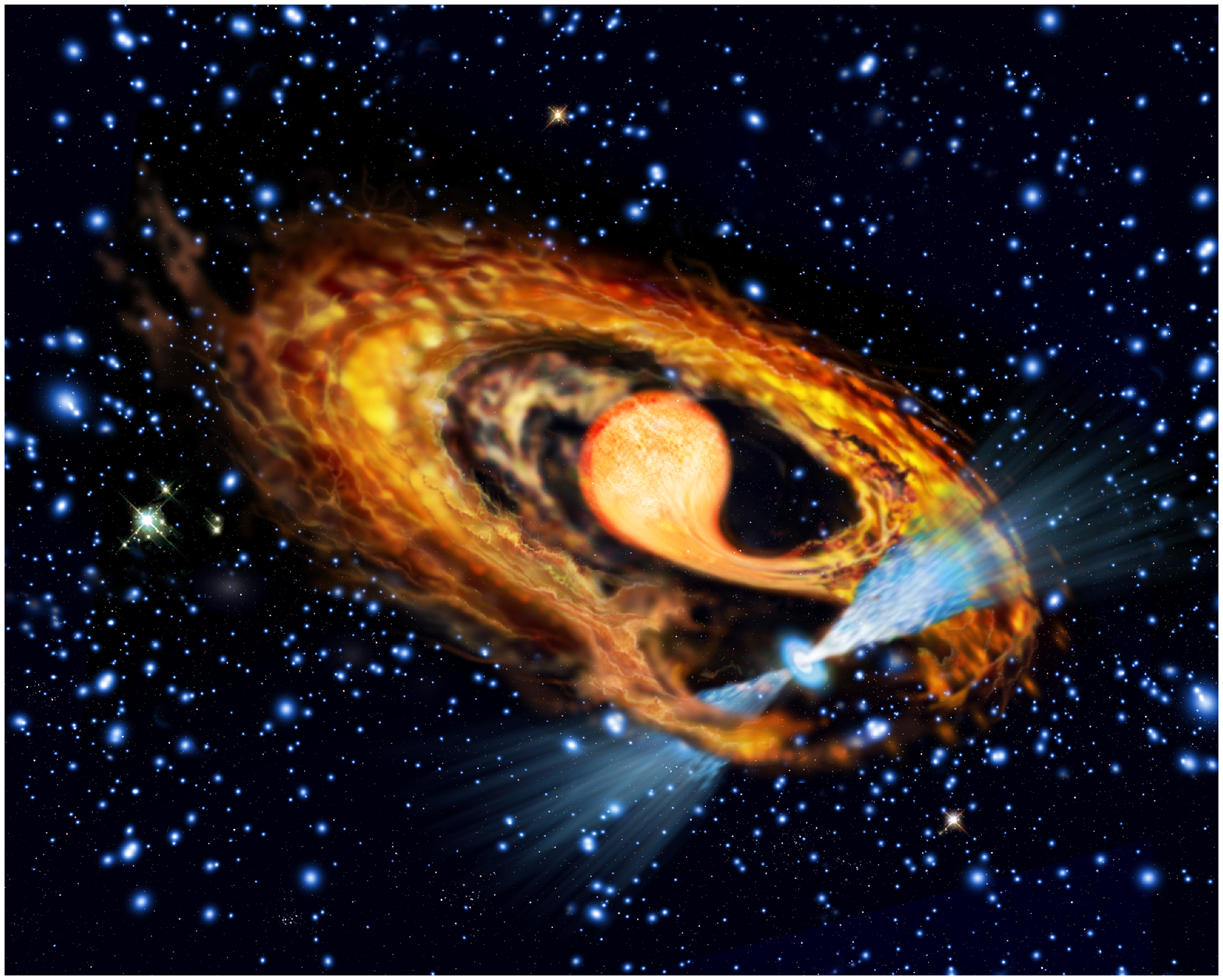}
  \label{fig:sfig1}
\end{subfigure}%
\begin{subfigure}{.5\textwidth}
  \centering
  \includegraphics[width=1.04\linewidth]{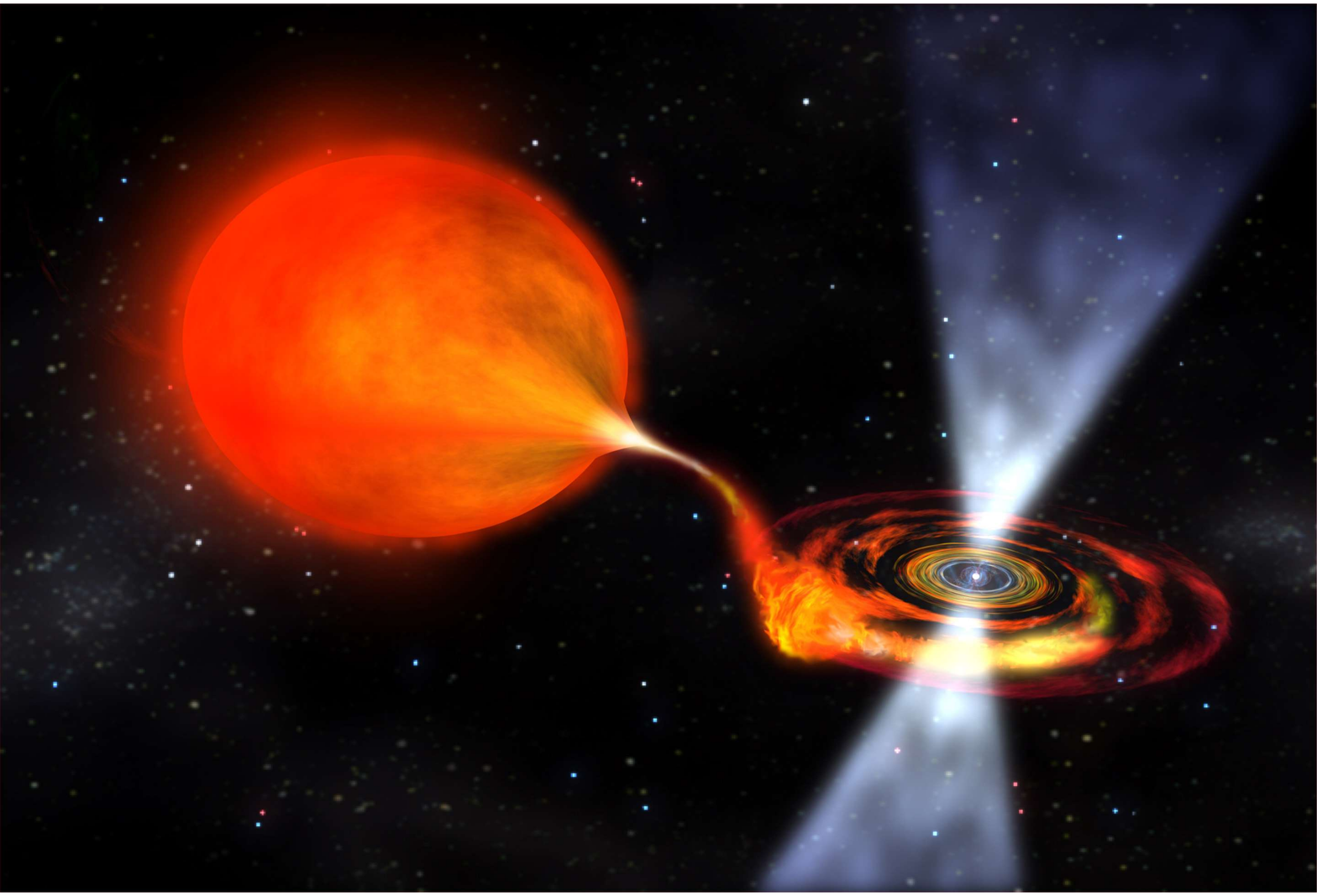}
  \label{fig:sfig2}
\end{subfigure}
\caption{Artistic impression of a system during the accretion phase (\textit{right panel}) and during a \textit{radio ejection}
phase \textit{left panel}. [Credits to NASA \& ESO, respectively]}
\label{fig:con_mt}
\end{figure}

Indeed, there are other indirect evidences of a non-conservative mass transfer in AMXPs. \cite{Marino2019} have analysed a sample of AMXPs, starting from XTE J0929-314 \cite{Marino2017}, finding that the averaged (over the time since their discovery) X-ray luminosity of most sources of the sample is significantly lower than what would be predicted by conservative mass transfer driven by GR and/or MB. Comparing their averaged X-ray luminosity with that predicted for a conservative mass transfer, a lower limit on the source distance may be estimated. Based on a sample of 18 sources, strong evidence of a non-conservative mass transfer was found for five sources, for which the estimated distance lower limits are significantly higher than their known distances, while hints of mass outflows are found in further six sources of the sample. The discrepancy can be fixed under the hypothesis of a non-conservative mass transfer in which a fraction of the mass transferred onto the compact object is swept away from the system, likely due to the radiation pressure of the rotating magnetic dipole and/or pulsar wind (see an artistic impression of the \textit{accretion} and \textit{ejection} phases in Fig.~\ref{fig:con_mt}). Interestingly, the possibility of strong outflows from these systems has been recently confirmed by general-relativistic MHD simulations \cite{2017ApJ...851L..34P} showing how the interaction of a turbulent accretion flow with the electromagnetic wind of the pulsar can lead to the transition of a rotational-powered pulsar to the accreting state, causing in turn the formation of relativistic jets whose power can greatly exceed the power of the pulsar wind. If the accretion rate is below a critical value, the pulsar may instead expel the accretion stream.

A similar argument has also been proposed for the black-hole X-ray Binary and microquasar V404 Cyg  \cite{Ziolkowski2018}; considering the donor evolution and mass transfer in the microquasar V404 Cyg and X-ray observations of its two outbursts, the authors find that the average mass accretion rate is substantially lower than the model mass-loss rate from the low-mass giant donor; to fix this discrepancy, they propose that a large fraction of the mass flowing from the donor leaves the binary in the form of outflows from the accretion disc around the accretor.   
We can conclude that, regardless of the nature of the accretor and the radiation emitted, it seems that radiation pressure has an important role in limiting the accretion of matter onto the compact object and in initiating a not-conservative mass transfer in LMXB systems, which may be a common feature in these systems, possibly related (as a cause or consequence) to the transient behavior itself.

\section{Summary and Open questions}
Despite the amount of information we have gained in the last two decades of observations and theoretical studies of AMXPs, several issue still remain to be addressed, as for instance the torque imparted by the accreting matter onto the NS, in most cases hidden inside the strong timing noise present in the timing of the spin period. Is the spin-up or spin-down of the NS overwhelmed with the large timing noise? Moreover, what is the origin of this large timing noise? Movements of the hot spot on the NS surface caused by flux variations have been proposed to explain the large timing noise, although it is not clear why in some sources (e.g. \saxj{}) is appears to be much stronger than in other sources (e.g. \ IGR~J00291+5934). Even more puzzling are the orbital residuals observed in \saxj{} and the different behaviour observed in IGR~J00291+5934, as well as the role of non-conservative mass transfer in AMXPs and LMXBs in general. Beside that, there are other important issues that should be addressed and are briefly described in the following. 

The discovery of AMXPs and the subsequent discovery of transitional millisecond pulsars (see Chapter 7 of this book for further details) has confirmed the recycling scenario. As a consequence of that, we improved our understanding of the formation of millisecond pulsars, which are accelerated by the accretion of matter and angular momentum during the LMXB phase, and of the evolutionary path linking the progenitors, i.e. LMXBs, to the end products of the evolution, i.e. Black-Widow pulsars and Redbacks, possibly through the transitional phase. Nevertheless, several open questions remain to be addressed, the first one regarding pulsations themselves.  
In fact, apart from the presence of coherent pulsations, AMXPs resemble the behaviour of transient LMXBs of the atoll class. Both the spectral properties and the aperiodic and quasi-periodic variability (so-called QPOs) are very similar between AMXPs and not-pulsating LMXBs (see e.g.\ \cite{Wijnands1999}, see also \cite{Patruno2012} for a review). Similar to LMXBs, AMXPs show type-I X-ray bursts and all the associated phenomenology, as for example the presence of (quasi-coherent) oscillations at the spin frequency of the NS during type-I X-ray bursts. From the observation of burst oscillations we know that many NS in LMXBs indeed rotate at millisecond periods. However, despite all these similarities, the large majority of LMXBs harbouring a NS do not show coherent pulsations, not even when the mass accretion rate decreases (for instance in transient systems) enough to allow the magnetosphere to expand outside the NS surface. The observation of an intermittent behaviour of coherent pulsations in some AMXPs (see e.g. the case of Aql X-1 or HETE J1900.1-2455) has suggested that magnetic field burial caused by accretion of fresh matter may play a role (see e.g. \cite{Cumming2001} and references therein). However, it is not clear whether this can explain the lack of pulsations in most of LMXBs or other factors contribute in hampering the detection of pulsations in these sources. These may be for instance a smearing of the pulsations by an optically thick corona, a smearing of pulsations due to gravitational light bending, alignment of the NS magnetic and rotational axes, onset of MHD instabilities at the disk/magnetospheric boundary. None of these models, however, furnish a satisfactorily explanation valid for all the cases (see a discussion in \cite{Patruno2012,Campana2018}).

Even more puzzling is the lack of radio pulsations in both AMXPs and LMXBs during X-ray quiescence. In principle, when the accretion of matter stops during (long) quiescent periods, the mechanism producing radio (or gamma-ray) pulsations should resume and the millisecond pulsar should shine in radio (or gamma-ray) as a rotation-powered pulsar. However, this has been observed to date in just one source, the AMXP and transitional pulsar IGR~J18245-2452 (J18245 hereafter) in the Globular Cluster M28 \cite{Papitto2013b}. This source has been first observed as a radio millisecond pulsar (spinning at 3.93 ms) in a binary system with a 11-h orbital period. In 2013 it went into X-ray outburst and was discovered as an AMXP; soon after the end of the outburst J18245 was detected again as a radio pulsar, demonstrating that the transition between the rotation-powered and the accretion-powered regime can occur on short timescales (in about 10 days or even less). It is worth noting that the other 2-3 sources, belonging to the transitional millisecond pulsar class, also show radio pulsations during X-ray quiescence and X-ray pulsations during the so-called disk state with (possibly) a low-level of accretion. However, none of these sources ever showed an X-ray outburst to date similar to the one showed by J18245 or the other AMXPs. The compelling possibility that these systems could swiftly switch from accretion-powered to rotation-powered magneto-dipole emitters during quiescence gives the opportunity to study a phase that could shed new light on the not yet cleared up radio pulsar emission mechanism. Therefore, if the swing between the rotation-powered and the accretion-powered pulsar can happen on fast timescales, why is this  observed just in few cases? Why radio millisecond pulsations at the known spin period have never been detected in other LMXBs or AMXPs during X-ray quiescence? 

In the framework of the so-called radio-ejection model \cite{Burderi2001}, the radio pulsar mechanism switches on when a significant reduction of the mass-transfer rate occurs. The accretion of matter onto the NS is then inhibited by the radiation pressure from the radio pulsar, which may be capable of ejecting out of the system most of the matter overflowing from the companion-star Roche lobe. One of the strongest predictions of this model is the presence, during the radio-ejection phase, of a strong wind of matter emanating from the system (see an artistic impression of a system in the \textit{radio-ejection} phase in the left panel of Fig.~\ref{fig:con_mt}). The non-detection of radio pulsations in this situation may be due to free-free absorption of the radio signal interacting with the ejected matter. A possibility to test this scenario is, therefore, to perform deep (tens of hours) radio observations of these sources at high radio frequency (above $5-6$ GHz), since the cross-section of free-free absorption decreases with frequency as $\nu^{-2}$ (see e.g. \cite{Iacolina2009,Iacolina2010}). However, the question remains: why do transitional millisecond pulsars, and J18245 in particular, behave in a different way, showing radio pulsations when the X-ray emission is off? Perhaps, a favourable geometry, e.g. a relatively low inclination angle of the system, may reduce the amount of matter along our line of sight, since most of the matter is expected to lie in the equatorial plane, and therefore reduce the amount of free-free absorption in these systems. Alternatively, pulsating radio emission should be searched in systems with long orbital periods, in which the matter transferred by the companion star is spread over a wide orbit. 

Despite the fact that radio pulsations remain elusive in AMXPs, sporadic detection of transient emission in the radio band has been reported in a few cases. On the other extreme of the electromagnetic spectrum, in the gamma-ray band, searches of AMXPs counterpart is also quite difficult. Because of the paucity of photons at such high energies, in order to obtain a statistically significant detection, it is necessary to integrate over several years. The analysis of $\sim 6$ yr of data from the Large Area Telescope on board the Fermi gamma-ray Space Telescope (Fermi-LAT) revealed a possible gamma-ray counterpart of \saxj{}, at a significance of $\sim 6 \sigma$, with a position compatible with that of the source within the $95\%$ confidence level \cite{deOnaWilhelmi2016}. However, the search for coherent pulsations did not produce a significant detection taking into account the number of trials. Uncertainties in the source position, orbital motion of the pulsar as well as the intrinsic evolution of the pulsar spin, which still are not known with enough precision to maintain the phase over years, are likely the causes of the non detection. A precise knowledge of the spin and orbital parameters of AMXPs is of fundamental importance to allow deep searches of their counterparts in the gamma-ray band, which has the advantage of not suffering the free-free absorption as in the radio band, but the disadvantage of the reduced number of photons, which requires folding over years in order to reach the statistics needed for detecting a (weak) pulsed signal. 

On other other hand, searches of the optical counterpart of these systems has given interesting, unexpected results. In several AMXPs, the optical counterpart during X-ray quiescence appears surprisingly luminous, inconsistent with both intrinsic emission from the companion star and X-ray reprocessing (e.g. \cite{Homer2001,DAvanzo2009,DAvanzo2011}). In fact, the optical counterpart shows an approximately sinusoidal modulation with photometric minimum at the superior conjunction of the pulsar. The lack of ellipsoidal, double-humped variations, rules out an origin from intrinsic emission from the companion star, while it is best explained as caused by the irradiated surface of the companion star. Given the lack of significant X-ray emission during quiescence, this has been interpreted as a strong (indirect) evidence that a rotating magneto-dipole powers the quiescent emission of AMXPs \cite{Burderi2003,Campana2004}. In fact, the magnetic dipole rotator, if active during quiescence, has a bolometric luminosity given by the Larmor's formula and may power the reprocessed optical emission. 

Even more puzzling is the recent discovery of optical pulsations at the NS spin period in one of the transitional pulsars, PSR J1023+0038 \cite{Ambrosino2017,Papitto2019}, the first time ever from a millisecond pulsar. Optical pulsations, with a maximum pulsed optical luminosity of $L_{pulse} \simeq 0.01 L_{opt} \simeq 10^{31}$ erg s$^{-1}$, were observed when the source was in a bright active state corresponding to an X-ray luminosity of $7\times 10^{33}$ erg s$^{-1}$ (see Chapter 7 of this book). 

From the discussion above it is clear that, despite the amount of observations and information obtained on AMXPs to date, there are still several issues that deserve further investigation, also considering that some new discoveries have raised other new questions. Nevertheless, one of the most important open questions about AMXPs is their spin period distribution and, most of all, the minimum spin period for a NS. Since (recycled) millisecond pulsars are accelerated during the LMXB phase, we expect that the minimum period of a NS is reached during this accretion phase, before the starting of the (non-accreting) spin-down phase caused by the emission of the magnetic dipole rotator. Hence, we expect that the fastest spinning NS should reside in an AMXP. Since the maximum frequency of a NS depends on its compactness, that is on its mass to radius ratio, the detection of the maximum spin frequency of NS may give strong and important constraints on the EoS of ultra-dense matter.
However the distribution of spin frequencies of the ensemble of AMXPs has an abrupt cutoff at about 730 Hz (e.g. \cite{Patruno2017b}, see also \cite{Papitto2014}), well below the maximum spin frequency allowed by the majority of realistic EoS. We are left therefore with the following questions: is the maximum frequency of NS telling us something related to the EoS of ultra-dense matter? Alternatively, which is the factor limiting the spin of NS well below the maximum possible possible frequency? Several possibilities have been proposed as a limiting factor for the NS rotation, such as emission of Gravitational Radiation \cite{Hartman2011,Papitto2011b}, the presence of a (not completely decayed) magnetic field \cite{ Patruno2012}, bias caused by a fast orbital motion \cite{Burderi2001}, and so on. However, none of these possibilities seems to explain all the phenomenology of AMXPs and LMXBs, and further investigation is needed to assess this fascinating question. To this aim, future X-ray missions, with large effective area and fast timing capabilities, such as {\it Athena}, possibly coupled with polarimetric capabilities, as is the case of the enhanced X-ray Timing and Polarimetry mission, {\it eXTP}, may be fundamental to put forward the research in this field and to open an new era of exciting discoveries on millisecond pulsars.

\bibliographystyle{spmpsci.bst}
\bibliography{export-bibtex.bib}

\begin{thebibliography}{100}
\providecommand{\url}[1]{{#1}}
\providecommand{\urlprefix}{URL }
\expandafter\ifx\csname urlstyle\endcsname\relax
  \providecommand{\doi}[1]{DOI~\discretionary{}{}{}#1}\else
  \providecommand{\doi}{DOI~\discretionary{}{}{}\begingroup
  \urlstyle{rm}\Url}\fi

\bibitem{Altamirano2008}
{Altamirano}, D., {Casella}, P., {Patruno}, A., {Wijnands}, R., {van der Klis},
  M.: {Intermittent Millisecond X-Ray Pulsations from the Neutron Star X-Ray
  Transient SAX J1748.9-2021 in the Globular Cluster NGC 6440}.
\newblock \apjl \textbf{674}(1), L45 (2008).
\newblock \doi{10.1086/528983}

\bibitem{Altamirano2011}
{Altamirano}, D., {Cavecchi}, Y., {Patruno}, A., {Watts}, A., {Linares}, M.,
  {Degenaar}, N., {Kalamkar}, M., {van der Klis}, M., {Rea}, N., {Casella}, P.,
  {Armas Padilla}, M., {Kaur}, R., {Yang}, Y.J., {Soleri}, P., {Wijnands}, R.:
  {Discovery of an Accreting Millisecond Pulsar in the Eclipsing Binary System
  SWIFT J1749.4-2807}.
\newblock \apjl \textbf{727}(1), L18 (2011).
\newblock \doi{10.1088/2041-8205/727/1/L18}

\bibitem{Altamirano2010}
{Altamirano}, D., {Patruno}, A., {Heinke}, C.O., {Markwardt}, C., {Strohmayer},
  T.E., {Linares}, M., {Wijnand s}, R., {van der Klis}, M., {Swank}, J.H.:
  {Discovery of a 205.89 Hz Accreting Millisecond X-ray Pulsar in the Globular
  Cluster NGC 6440}.
\newblock \apjl \textbf{712}(1), L58--L62 (2010).
\newblock \doi{10.1088/2041-8205/712/1/L58}

\bibitem{Ambrosino2017}
{Ambrosino}, F., {Papitto}, A., {Stella}, L., {Meddi}, F., {Cretaro}, P.,
  {Burderi}, L., {Di Salvo}, T., {Israel}, G.L., {Ghedina}, A., {Di Fabrizio},
  L., {Riverol}, L.: {Optical pulsations from a transitional millisecond
  pulsar}.
\newblock Nature Astronomy \textbf{1}, 854--858 (2017).
\newblock \doi{10.1038/s41550-017-0266-2}

\bibitem{Applegate1992}
{Applegate}, J.H.: {A Mechanism for Orbital Period Modulation in Close
  Binaries}.
\newblock \apj \textbf{385}, 621 (1992).
\newblock \doi{10.1086/170967}

\bibitem{Applegate1994}
{Applegate}, J.H., {Shaham}, J.: {Orbital Period Variability in the Eclipsing
  Pulsar Binary PSR B1957+20: Evidence for a Tidally Powered Star}.
\newblock \apj \textbf{436}, 312 (1994).
\newblock \doi{10.1086/174906}

\bibitem{Benvenuto2017}
{Benvenuto}, O.G., {De Vito}, M.A., {Horvath}, J.E.: {Evolution of redback
  radio pulsars in globular clusters}.
\newblock \aap \textbf{598}, A35 (2017).
\newblock \doi{10.1051/0004-6361/201628692}

\bibitem{Bhattacharya1991}
{Bhattacharya}, D., {van den Heuvel}, E.P.J.: {Formation and evolution of
  binary and millisecond radio pulsars}.
\newblock \physrep \textbf{203}(1-2), 1--124 (1991).
\newblock \doi{10.1016/0370-1573(91)90064-S}

\bibitem{Bildsten2001}
{Bildsten}, L., {Chakrabarty}, D.: {A Brown Dwarf Companion for the Accreting
  Millisecond Pulsar SAX J1808.4-3658}.
\newblock \apj \textbf{557}(1), 292--296 (2001).
\newblock \doi{10.1086/321633}

\bibitem{Bult2018b}
{Bult}, P., {Altamirano}, D., {Arzoumanian}, Z., {Chakrabarty}, D., {Gendreau},
  K.C., {Guillot}, S., {Ho}, W.C.G., {Jaisawal}, G.K., {Lentine}, S.,
  {Markwardt}, C.B., {Ngo}, S.N., {Pope}, J.S., {Ray}, P.S., {Saylor}, M.R.,
  {Strohmayer}, T.E.: {On the 2018 Outburst of the Accreting Millisecond X-Ray
  Pulsar Swift J1756.9-2508 As Seen with NICER}.
\newblock \apj \textbf{864}(1), 14 (2018).
\newblock \doi{10.3847/1538-4357/aad5e5}

\bibitem{Bult2020}
{Bult}, P., {Chakrabarty}, D., {Arzoumanian}, Z., {Gendreau}, K.C., {Guillot},
  S., {Malacaria}, C., {Ray}, P.S., {Strohmayer}, T.E.: {Timing the Pulsations
  of the Accreting Millisecond Pulsar SAX J1808.4-3658 during Its 2019
  Outburst}.
\newblock \apj \textbf{898}(1), 38 (2020).
\newblock \doi{10.3847/1538-4357/ab9827}

\bibitem{Bult2015c}
{Bult}, P., {Patruno}, A., {van der Klis}, M.: {Coherent Timing of the
  Accreting Millisecond Pulsar NGC 6440 X-2}.
\newblock \apj \textbf{814}(2), 138 (2015).
\newblock \doi{10.1088/0004-637X/814/2/138}

\bibitem{Burderi2003}
{Burderi}, L., {Di Salvo}, T., {D'Antona}, F., {Robba}, N.R., {Testa}, V.: {The
  optical counterpart to SAX J1808.4-3658 in quiescence: Evidence of an active
  radio pulsar?}
\newblock \aap \textbf{404}, L43--L46 (2003).
\newblock \doi{10.1051/0004-6361:20030669}

\bibitem{Burderi2007}
{Burderi}, L., {Di Salvo}, T., {Lavagetto}, G., {Menna}, M.T., {Papitto}, A.,
  {Riggio}, A., {Iaria}, R., {D'Antona}, F., {Robba}, N.R., {Stella}, L.:
  {Timing an Accreting Millisecond Pulsar: Measuring the Accretion Torque in
  IGR J00291+5934}.
\newblock \apj \textbf{657}(2), 961--966 (2007).
\newblock \doi{10.1086/510659}

\bibitem{Burderi2006}
{Burderi}, L., {Di Salvo}, T., {Menna}, M.T., {Riggio}, A., {Papitto}, A.:
  {Order in the Chaos: Spin-up and Spin-down during the 2002 Outburst of SAX
  J1808.4-3658}.
\newblock \apjl \textbf{653}(2), L133--L136 (2006).
\newblock \doi{10.1086/510666}

\bibitem{Burderi2010}
{Burderi}, L., {Di Salvo}, T., {Riggio}, A., {Papitto}, A., {Iaria}, R.,
  {D'A{\`\i}}, A., {Menna}, M.T.: {New ephemeris of the ADC source 2A 1822-371:
  a stable orbital-period derivative over 30 years}.
\newblock \aap \textbf{515}, A44 (2010).
\newblock \doi{10.1051/0004-6361/200912881}

\bibitem{Burderi1998}
{Burderi}, L., {King}, A.R.: {A Firm Upper Limit to the Radius of the Neutron
  Star in SAX J1808.4-3658}.
\newblock \apjl \textbf{505}(2), L135--L137 (1998).
\newblock \doi{10.1086/311611}

\bibitem{Burderi1999}
{Burderi}, L., {Possenti}, A., {Colpi}, M., {Di Salvo}, T., {D'Amico}, N.:
  {Neutron Stars with Submillisecond Periods: A Population of High-Mass
  Objects?}
\newblock \apj \textbf{519}(1), 285--290 (1999).
\newblock \doi{10.1086/307353}

\bibitem{Burderi2001}
{Burderi}, L., {Possenti}, A., {D'Antona}, F., {Di Salvo}, T., {Burgay}, M.,
  {Stella}, L., {Menna}, M.T., {Iaria}, R., {Campana}, S., {d'Amico}, N.:
  {Where May Ultrafast Rotating Neutron Stars Be Hidden?}
\newblock \apjl \textbf{560}(1), L71--L74 (2001).
\newblock \doi{10.1086/324220}

\bibitem{Burderi2009}
{Burderi}, L., {Riggio}, A., {di Salvo}, T., {Papitto}, A., {Menna}, M.T.,
  {D'A{\`\i}}, A., {Iaria}, R.: {Timing of the 2008 outburst of SAX
  J1808.4-3658 with XMM-Newton: a stable orbital-period derivative over ten
  years}.
\newblock \aap \textbf{496}(2), L17--L20 (2009).
\newblock \doi{10.1051/0004-6361/200811542}

\bibitem{Cackett2009}
{Cackett}, E.M., {Altamirano}, D., {Patruno}, A., {Miller}, J.M., {Reynolds},
  M., {Linares}, M., {Wijnands}, R.: {Broad Relativistic Iron Emission Line
  Observed in SAX J1808.4-3658}.
\newblock \apjl \textbf{694}(1), L21--L25 (2009).
\newblock \doi{10.1088/0004-637X/694/1/L21}

\bibitem{Cadelano2017}
{Cadelano}, M., {Pallanca}, C., {Ferraro}, F.R., {Dalessand ro}, E., {Lanzoni},
  B., {Patruno}, A.: {The Optical Counterpart to the Accreting Millisecond
  X-Ray Pulsar SAX J1748.9-2021 in the Globular Cluster NGC 6440}.
\newblock \apj \textbf{844}(1), 53 (2017).
\newblock \doi{10.3847/1538-4357/aa7b7f}

\bibitem{Campana2013}
{Campana}, S., {Coti Zelati}, F., {D'Avanzo}, P.: {Mining the Aql X-1 long-term
  X-ray light curve}.
\newblock \mnras \textbf{432}(2), 1695--1700 (2013).
\newblock \doi{10.1093/mnras/stt604}

\bibitem{Campana2004}
{Campana}, S., {D'Avanzo}, P., {Casares}, J., {Covino}, S., {Israel}, G.,
  {Marconi}, G., {Hynes}, R., {Charles}, P., {Stella}, L.: {Indirect Evidence
  of an Active Radio Pulsar in the Quiescent State of the Transient Millisecond
  Pulsar SAX J1808.4-3658}.
\newblock \apjl \textbf{614}(1), L49--L52 (2004).
\newblock \doi{10.1086/425495}

\bibitem{Campana2018}
{Campana}, S., {Di Salvo}, T.: {Accreting Pulsars: Mixing-up Accretion Phases
  in Transitional Systems}, \emph{Astrophysics and Space Science Library}, vol.
  457, p. 149 (2018).
\newblock \doi{10.1007/978-3-319-97616-7-4}

\bibitem{Campana2003}
{Campana}, S., {Ravasio}, M., {Israel}, G.L., {Mangano}, V., {Belloni}, T.:
  {XMM-Newton Observation of the 5.25 Millisecond Transient Pulsar XTE
  J1807-294 in Outburst}.
\newblock \apjl \textbf{594}(1), L39--L42 (2003).
\newblock \doi{10.1086/378258}

\bibitem{Casella2008}
{Casella}, P., {Altamirano}, D., {Patruno}, A., {Wijnands}, R., {van der Klis},
  M.: {Discovery of Coherent Millisecond X-Ray Pulsations in Aquila X-1}.
\newblock \apjl \textbf{674}(1), L41 (2008).
\newblock \doi{10.1086/528982}

\bibitem{Chakrabarty1998}
{Chakrabarty}, D., {Morgan}, E.H.: {The two-hour orbit of a binary millisecond
  X-ray pulsar}.
\newblock \nat \textbf{394}(6691), 346--348 (1998).
\newblock \doi{10.1038/28561}

\bibitem{Chen2017}
{Chen}, W.C.: {An evolutionary channel towards the accreting millisecond pulsar
  SAX J1808.4-3658}.
\newblock \mnras \textbf{464}(4), 4673--4679 (2017).
\newblock \doi{10.1093/mnras/stw2747}

\bibitem{Chou2016}
{Chou}, Y., {Hsieh}, H.E., {Hu}, C.P., {Yang}, T.C., {Su}, Y.H.: {Orbital and
  Spin Parameter Variations of Partial Eclipsing Low Mass X-Ray Binary X
  1822-371}.
\newblock \apj \textbf{831}(1), 29 (2016).
\newblock \doi{10.3847/0004-637X/831/1/29}

\bibitem{Cornelisse2009}
{Cornelisse}, R., {D'Avanzo}, P., {Mu{\~n}oz-Darias}, T., {Campana}, S.,
  {Casares}, J., {Charles}, P.A., {Steeghs}, D., {Israel}, G., {Stella}, L.:
  {Phase-resolved spectroscopy of the accreting millisecond X-ray pulsar SAX
  J1808.4-3658 during the 2008 outburst}.
\newblock \aap \textbf{495}(1), L1--L4 (2009).
\newblock \doi{10.1051/0004-6361:200811396}

\bibitem{Cumming2001}
{Cumming}, A., {Zweibel}, E., {Bildsten}, L.: {Magnetic Screening in Accreting
  Neutron Stars}.
\newblock \apj \textbf{557}(2), 958--966 (2001).
\newblock \doi{10.1086/321658}

\bibitem{DAvanzo2009}
{D'Avanzo}, P., {Campana}, S., {Casares}, J., {Covino}, S., {Israel}, G.L.,
  {Stella}, L.: {The optical counterparts of accreting millisecond X-ray
  pulsars during quiescence}.
\newblock \aap \textbf{508}(1), 297--308 (2009).
\newblock \doi{10.1051/0004-6361/200810249}

\bibitem{DAvanzo2011}
{D'Avanzo}, P., {Campana}, S., {Mu{\~n}oz-Darias}, T., {Belloni}, T., {Bozzo},
  E., {Falanga}, M., {Stella}, L.: {A search for the near-infrared counterpart
  of the eclipsing millisecond X-ray pulsar Swift J1749.4-2807}.
\newblock \aap \textbf{534}, A92 (2011).
\newblock \doi{10.1051/0004-6361/201117841}

\bibitem{deOnaWilhelmi2016}
{de O{\~n}a Wilhelmi}, E., {Papitto}, A., {Li}, J., {Rea}, N., {Torres}, D.F.,
  {Burderi}, L., {Di Salvo}, T., {Iaria}, R., {Riggio}, A., {Sanna}, A.: {SAX
  J1808.4-3658, an accreting millisecond pulsar shining in gamma rays?}
\newblock \mnras \textbf{456}(3), 2647--2653 (2016).
\newblock \doi{10.1093/mnras/stv2695}

\bibitem{Degenaar2017b}
{Degenaar}, N., {Ootes}, L.S., {Reynolds}, M.T., {Wijnand s}, R., {Page}, D.:
  {A cold neutron star in the transient low-mass X-ray binary HETE J1900.1-2455
  after 10 yr of active accretion}.
\newblock \mnras \textbf{465}(1), L10--L14 (2017).
\newblock \doi{10.1093/mnrasl/slw197}

\bibitem{Deloye2008}
{Deloye}, C.J., {Heinke}, C.O., {Taam}, R.E., {Jonker}, P.G.: {Optical
  observations of SAX J1808.4-3658 during quiescence}.
\newblock \mnras \textbf{391}(4), 1619--1628 (2008).
\newblock \doi{10.1111/j.1365-2966.2008.14021.x}

\bibitem{DiSalvo2008}
{di Salvo}, T., {Burderi}, L., {Riggio}, A., {Papitto}, A., {Menna}, M.T.:
  {Orbital evolution of an accreting millisecond pulsar: witnessing the banquet
  of a hidden black widow?}
\newblock \mnras \textbf{389}(4), 1851--1857 (2008).
\newblock \doi{10.1111/j.1365-2966.2008.13709.x}

\bibitem{DiSalvo2000}
{Di Salvo}, T., {Iaria}, R., {Burderi}, L., {Robba}, N.R.: {The Broadband
  Spectrum of MXB 1728-34 Observed by BeppoSAX}.
\newblock \apj \textbf{542}(2), 1034--1040 (2000).
\newblock \doi{10.1086/317029}

\bibitem{DiSalvo2019}
{Di Salvo}, T., {Sanna}, A., {Burderi}, L., {Papitto}, A., {Iaria}, R.,
  {Gambino}, A.F., {Riggio}, A.: {NuSTAR and XMM-Newton broad-band spectrum of
  SAX J1808.4-3658 during its latest outburst in 2015}.
\newblock \mnras \textbf{483}(1), 767--779 (2019).
\newblock \doi{10.1093/mnras/sty2974}

\bibitem{Done2007}
{Done}, C., {Gierli{\'n}ski}, M., {Kubota}, A.: {Modelling the behaviour of
  accretion flows in X-ray binaries. Everything you always wanted to know about
  accretion but were afraid to ask}.
\newblock \aapr \textbf{15}(1), 1--66 (2007).
\newblock \doi{10.1007/s00159-007-0006-1}

\bibitem{Elebert2008}
{Elebert}, P., {Callanan}, P.J., {Filippenko}, A.V., {Garnavich}, P.M.,
  {Mackie}, G., {Hill}, J.M., {Burwitz}, V.: {Optical photometry and
  spectroscopy of the accretion-powered millisecond pulsar HETE J1900.1 -
  2455}.
\newblock \mnras \textbf{383}(4), 1581--1587 (2008).
\newblock \doi{10.1111/j.1365-2966.2007.12667.x}

\bibitem{Falanga2005a}
{Falanga}, M., {Bonnet-Bidaud}, J.M., {Poutanen}, J., {Farinelli}, R.,
  {Martocchia}, A., {Goldoni}, P., {Qu}, J.L., {Kuiper}, L., {Goldwurm}, A.:
  {INTEGRAL spectroscopy of the accreting millisecond pulsar XTE J1807-294 in
  outburst}.
\newblock \aap \textbf{436}(2), 647--652 (2005).
\newblock \doi{10.1051/0004-6361:20042575}

\bibitem{Falanga2005b}
{Falanga}, M., {Kuiper}, L., {Poutanen}, J., {Bonning}, E.W., {Hermsen}, W.,
  {di Salvo}, T., {Goldoni}, P., {Goldwurm}, A., {Shaw}, S.E., {Stella}, L.:
  {INTEGRAL and RXTE observations of accreting millisecond pulsar IGR
  J00291+5934 in outburst}.
\newblock \aap \textbf{444}(1), 15--24 (2005).
\newblock \doi{10.1051/0004-6361:20053472}

\bibitem{Galloway2002}
{Galloway}, D.K., {Chakrabarty}, D., {Morgan}, E.H., {Remillard}, R.A.:
  {Discovery of a High-Latitude Accreting Millisecond Pulsar in an Ultracompact
  Binary}.
\newblock \apjl \textbf{576}(2), L137--L140 (2002).
\newblock \doi{10.1086/343841}

\bibitem{Galloway2005}
{Galloway}, D.K., {Markwardt}, C.B., {Morgan}, E.H., {Chakrabarty}, D.,
  {Strohmayer}, T.E.: {Discovery of the Accretion-powered Millisecond X-Ray
  Pulsar IGR J00291+5934}.
\newblock \apjl \textbf{622}(1), L45--L48 (2005).
\newblock \doi{10.1086/429563}

\bibitem{Ghosh1979}
{Ghosh}, P., {Lamb}, F.K.: {Accretion by rotating magnetic neutron stars. III.
  Accretion torques and period changes in pulsating X-ray sources.}
\newblock \apj \textbf{234}, 296--316 (1979).
\newblock \doi{10.1086/157498}

\bibitem{Ghosh1991}
{Ghosh}, P., {Lamb}, F.K.: {Plasma Physics of Accreting Neutron Stars}.
\newblock In: J.~{Ventura}, D.~{Pines} (eds.) NATO Advanced Science Institutes
  (ASI) Series C, \emph{NATO Advanced Science Institutes (ASI) Series C}, vol.
  344, p. 363 (1991)

\bibitem{Gierlinski2002}
{Gierli{\'n}ski}, M., {Done}, C., {Barret}, D.: {Phase-resolved X-ray
  spectroscopy of the millisecond pulsar SAX J1808.4-3658}.
\newblock \mnras \textbf{331}(1), 141--153 (2002).
\newblock \doi{10.1046/j.1365-8711.2002.05174.x}

\bibitem{Gierlinski2005}
{Gierli{\'n}ski}, M., {Poutanen}, J.: {Physics of accretion in the millisecond
  pulsar XTE J1751-305}.
\newblock \mnras \textbf{359}(4), 1261--1276 (2005).
\newblock \doi{10.1111/j.1365-2966.2005.09004.x}

\bibitem{Giles2005}
{Giles}, A.B., {Greenhill}, J.G., {Hill}, K.M., {Sand ers}, E.: {The optical
  counterpart of XTE J0929-314: the third transient millisecond X-ray pulsar}.
\newblock \mnras \textbf{361}(4), 1180--1186 (2005).
\newblock \doi{10.1111/j.1365-2966.2005.09255.x}

\bibitem{Hartman2011}
{Hartman}, J.M., {Galloway}, D.K., {Chakrabarty}, D.: {A Double Outburst from
  IGR J00291+5934: Implications for Accretion Disk Instability Theory}.
\newblock \apj \textbf{726}(1), 26 (2011).
\newblock \doi{10.1088/0004-637X/726/1/26}

\bibitem{Hartman2008}
{Hartman}, J.M., {Patruno}, A., {Chakrabarty}, D., {Kaplan}, D.L., {Markwardt},
  C.B., {Morgan}, E.H., {Ray}, P.S., {van der Klis}, M., {Wijnands}, R.: {The
  Long-Term Evolution of the Spin, Pulse Shape, and Orbit of the
  Accretion-powered Millisecond Pulsar SAX J1808.4-3658}.
\newblock \apj \textbf{675}(2), 1468--1486 (2008).
\newblock \doi{10.1086/527461}

\bibitem{Hartman2009b}
{Hartman}, J.M., {Patruno}, A., {Chakrabarty}, D., {Markwardt}, C.B., {Morgan},
  E.H., {van der Klis}, M., {Wijnands}, R.: {A Decade of Timing an
  Accretion-powered Millisecond Pulsar: The Continuing Spin Down and Orbital
  Evolution of SAX J1808.4-3658}.
\newblock \apj \textbf{702}(2), 1673--1678 (2009).
\newblock \doi{10.1088/0004-637X/702/2/1673}

\bibitem{Hartman2009a}
{Hartman}, J.M., {Watts}, A.L., {Chakrabarty}, D.: {The Luminosity and Energy
  Dependence of Pulse Phase Lags in the Accretion-powered Millisecond Pulsar
  SAX J1808.4-3658}.
\newblock \apj \textbf{697}(2), 2102--2107 (2009).
\newblock \doi{10.1088/0004-637X/697/2/2102}

\bibitem{Haskell2018}
{Haskell}, B., {Zdunik}, J.L., {Fortin}, M., {Bejger}, M., {Wijnands}, R.,
  {Patruno}, A.: {Fundamental physics and the absence of sub-millisecond
  pulsars}.
\newblock \aap \textbf{620}, A69 (2018).
\newblock \doi{10.1051/0004-6361/201833521}

\bibitem{Hessels2006}
{Hessels}, J.W.T., {Ransom}, S.M., {Stairs}, I.H., {Freire}, P.C.C., {Kaspi},
  V.M., {Camilo}, F.: {A Radio Pulsar Spinning at 716 Hz}.
\newblock Science \textbf{311}(5769), 1901--1904 (2006).
\newblock \doi{10.1126/science.1123430}

\bibitem{Homer2001}
{Homer}, L., {Charles}, P.A., {Chakrabarty}, D., {van Zyl}, L.: {The optical
  counterpart to SAX J1808.4-3658: observations in quiescence}.
\newblock \mnras \textbf{325}(4), 1471--1476 (2001).
\newblock \doi{10.1046/j.1365-8711.2001.04567.x}

\bibitem{Iacolina2009}
{Iacolina}, M.N., {Burgay}, M., {Burderi}, L., {Possenti}, A., {di Salvo}, T.:
  {Searching for pulsed emission from XTE J0929-314 at high radio frequencies}.
\newblock \aap \textbf{497}(2), 445--450 (2009).
\newblock \doi{10.1051/0004-6361/200810677}

\bibitem{Iacolina2010}
{Iacolina}, M.N., {Burgay}, M., {Burderi}, L., {Possenti}, A., {di Salvo}, T.:
  {Search for pulsations at high radio frequencies from accreting millisecond
  X-ray pulsars in quiescence}.
\newblock \aap \textbf{519}, A13 (2010).
\newblock \doi{10.1051/0004-6361/201014025}

\bibitem{Iaria2011}
{Iaria}, R., {di Salvo}, T., {Burderi}, L., {D'A{\'\i}}, A., {Papitto}, A.,
  {Riggio}, A., {Robba}, N.R.: {Detailed study of the X-ray and optical/UV
  orbital ephemeris of X1822-371}.
\newblock \aap \textbf{534}, A85 (2011).
\newblock \doi{10.1051/0004-6361/201117334}

\bibitem{Ibragimov2009}
{Ibragimov}, A., {Poutanen}, J.: {Accreting millisecond pulsar SAX J1808.4-3658
  during its 2002 outburst: evidence for a receding disc}.
\newblock \mnras \textbf{400}(1), 492--508 (2009).
\newblock \doi{10.1111/j.1365-2966.2009.15477.x}

\bibitem{Jackson}
Jackson, J.D.: {Classical electrodynamics; 2nd ed.}
\newblock Wiley, New York, NY (1975).
\newblock \urlprefix\url{https://cds.cern.ch/record/100964}

\bibitem{Jonker2001}
{Jonker}, P.G., {van der Klis}, M.: {Discovery of an X-Ray Pulsar in the
  Low-Mass X-Ray Binary 2A 1822-371}.
\newblock \apjl \textbf{553}(1), L43--L46 (2001).
\newblock \doi{10.1086/320510}

\bibitem{Kaaret2006}
{Kaaret}, P., {Morgan}, E.H., {Vanderspek}, R., {Tomsick}, J.A.: {Discovery of
  the Millisecond X-Ray Pulsar HETE J1900.1-2455}.
\newblock \apj \textbf{638}(2), 963--967 (2006).
\newblock \doi{10.1086/498886}

\bibitem{Kluzniak2007}
{Klu{\'z}niak}, W., {Rappaport}, S.: {Magnetically Torqued Thin Accretion
  Disks}.
\newblock \apj \textbf{671}(2), 1990--2005 (2007).
\newblock \doi{10.1086/522954}

\bibitem{Krimm2007}
{Krimm}, H.A., {Markwardt}, C.B., {Deloye}, C.J., {Romano}, P., {Chakrabarty},
  D., {Campana}, S., {Cummings}, J.R., {Galloway}, D.K., {Gehrels}, N.,
  {Hartman}, J.M., {Kaaret}, P., {Morgan}, E.H., {Tueller}, J.: {Discovery of
  the Accretion-powered Millisecond Pulsar SWIFT J1756.9-2508 with a Low-Mass
  Companion}.
\newblock \apjl \textbf{668}(2), L147--L150 (2007).
\newblock \doi{10.1086/522959}

\bibitem{Kuiper2020}
{Kuiper}, L., {Tsygankov}, S.S., {Falanga}, M., {Mereminskij}, I.A.,
  {Galloway}, D.K., {Poutanen}, J., {Li}, Z.: {High-energy characteristics of
  the accretion-powered millisecond pulsar IGR J17591-2342 during its 2018
  outburst}.
\newblock arXiv e-prints arXiv:2002.12154 (2020)

\bibitem{2019A&A...627A.125M}
{Marino}, A., {Di Salvo}, T., {Burderi}, L., {Sanna}, A., {Riggio}, A.,
  {Papitto}, A., {Del Santo}, M., {Gambino}, A.F., {Iaria}, R., {Mazzola},
  S.M.: {Indications of non-conservative mass transfer in AMXPs}.
\newblock \aap \textbf{627}, A125 (2019).
\newblock \doi{10.1051/0004-6361/201834460}

\bibitem{Marino2019}
{Marino}, A., {Di Salvo}, T., {Burderi}, L., {Sanna}, A., {Riggio}, A.,
  {Papitto}, A., {Del Santo}, M., {Gambino}, A.F., {Iaria}, R., {Mazzola},
  S.M.: {Indications of non-conservative mass transfer in AMXPs}.
\newblock \aap \textbf{627}, A125 (2019).
\newblock \doi{10.1051/0004-6361/201834460}

\bibitem{Marino2017}
{Marino}, A., {Di Salvo}, T., {Gambino}, A.F., {Iaria}, R., {Burderi}, L.,
  {Matranga}, M., {Sanna}, A., {Riggio}, A.: {Evidence of a non-conservative
  mass transfer for XTE J0929-314}.
\newblock \aap \textbf{603}, A137 (2017).
\newblock \doi{10.1051/0004-6361/201730464}

\bibitem{Markwardt2003}
{Markwardt}, C.B., {Swank}, J.H.: {XTE J1814-338}.
\newblock \iaucirc \textbf{8144}, 1 (2003)

\bibitem{Markwardt2002}
{Markwardt}, C.B., {Swank}, J.H., {Strohmayer}, T.E., {in 't Zand}, J.J.M.,
  {Marshall}, F.E.: {Discovery of a Second Millisecond Accreting Pulsar: XTE
  J1751-305}.
\newblock \apjl \textbf{575}(1), L21--L24 (2002).
\newblock \doi{10.1086/342612}

\bibitem{MataSanchez2017}
{Mata S{\'a}nchez}, D., {Mu{\~n}oz-Darias}, T., {Casares}, J.,
  {Jim{\'e}nez-Ibarra}, F.: {The donor of Aquila X-1 revealed by high-angular
  resolution near-infrared spectroscopy}.
\newblock \mnras \textbf{464}(1), L41--L45 (2017).
\newblock \doi{10.1093/mnrasl/slw172}

\bibitem{Mazzola2019}
{Mazzola}, S.M., {Iaria}, R., {Di Salvo}, T., {Gambino}, A.F., {Marino}, A.,
  {Burderi}, L., {Sanna}, A., {Riggio}, A., {Tailo}, M.: {Updated orbital
  ephemeris of the ADC source X 1822-371: a stable orbital expansion over 40
  years}.
\newblock \aap \textbf{625}, L12 (2019).
\newblock \doi{10.1051/0004-6361/201935665}

\bibitem{Miller2003}
{Miller}, J.M., {Wijnands}, R., {M{\'e}ndez}, M., {Kendziorra}, E., {Tiengo},
  A., {van der Klis}, M., {Chakrabarty}, D., {Gaensler}, B.M., {Lewin}, W.H.G.:
  {XMM-Newton Spectroscopy of the Accretion-driven Millisecond X-Ray Pulsar XTE
  J1751-305 in Outburst}.
\newblock \apjl \textbf{583}(2), L99--L102 (2003).
\newblock \doi{10.1086/368105}

\bibitem{Mukherjee2015}
{Mukherjee}, D., {Bult}, P., {van der Klis}, M., {Bhattacharya}, D.: {The
  magnetic-field strengths of accreting millisecond pulsars}.
\newblock \mnras \textbf{452}(4), 3994--4012 (2015).
\newblock \doi{10.1093/mnras/stv1542}

\bibitem{Ozel2016}
{{\"O}zel}, F., {Freire}, P.: {Masses, Radii, and the Equation of State of
  Neutron Stars}.
\newblock \araa \textbf{54}, 401--440 (2016).
\newblock \doi{10.1146/annurev-astro-081915-023322}

\bibitem{Papitto2019}
{Papitto}, A., {Ambrosino}, F., {Stella}, L., {Torres}, D., {Coti Zelati}, F.,
  {Ghedina}, A., {Meddi}, F., {Sanna}, A., {Casella}, P., {Dallilar}, Y.,
  {Eikenberry}, S., {Israel}, G.L., {Onori}, F., {Piranomonte}, S., {Bozzo},
  E., {Burderi}, L., {Campana}, S., {de Martino}, D., {Di Salvo}, T.,
  {Ferrigno}, C., {Rea}, N., {Riggio}, A., {Serrano}, S., {Veledina}, A.,
  {Zampieri}, L.: {Pulsating in Unison at Optical and X-Ray Energies:
  Simultaneous High Time Resolution Observations of the Transitional
  Millisecond Pulsar PSR J1023+0038}.
\newblock \apj \textbf{882}(2), 104 (2019).
\newblock \doi{10.3847/1538-4357/ab2fdf}

\bibitem{Papitto2011b}
{Papitto}, A., {Bozzo}, E., {Ferrigno}, C., {Belloni}, T., {Burderi}, L., {di
  Salvo}, T., {Riggio}, A., {D'A{\`\i}}, A., {Iaria}, R.: {The discovery of the
  401 Hz accreting millisecond pulsar IGR J17498-2921 in a 3.8 h orbit}.
\newblock \aap \textbf{535}, L4 (2011).
\newblock \doi{10.1051/0004-6361/201117995}

\bibitem{Papitto2016}
{Papitto}, A., {Bozzo}, E., {Sanchez-Fernandez}, C., {Romano}, P., {Torres},
  D.F., {Ferrigno}, C., {Kajava}, J.J.E., {Kuulkers}, E.: {The 2015 outburst of
  the accreting millisecond pulsar IGR J17511-3057 as seen by INTEGRAL, Swift,
  and XMM-Newton}.
\newblock \aap \textbf{596}, A71 (2016).
\newblock \doi{10.1051/0004-6361/201628238}

\bibitem{Papitto2013a}
{Papitto}, A., {D'A{\`\i}}, A., {Di Salvo}, T., {Egron}, E., {Bozzo}, E.,
  {Burderi}, L., {Iaria}, R., {Riggio}, A., {Menna}, M.T.: {The accretion flow
  to the intermittent accreting millisecond pulsar, HETE J1900.1-2455, as
  observed by XMM-Newton and RXTE}.
\newblock \mnras \textbf{429}(4), 3411--3422 (2013).
\newblock \doi{10.1093/mnras/sts605}

\bibitem{Papitto2012}
{Papitto}, A., {Di Salvo}, T., {Burderi}, L., {Belloni}, T.M., {Stella}, L.,
  {Bozzo}, E., {D'A{\`\i}}, A., {Ferrigno}, C., {Iaria}, R., {Motta}, S.,
  {Riggio}, A., {Tramacere}, A.: {The pulse profile and spin evolution of the
  accreting pulsar in Terzan 5, IGR J17480-2446, during its 2010 outburst}.
\newblock \mnras \textbf{423}(2), 1178--1193 (2012).
\newblock \doi{10.1111/j.1365-2966.2012.20945.x}

\bibitem{Papitto2007}
{Papitto}, A., {di Salvo}, T., {Burderi}, L., {Menna}, M.T., {Lavagetto}, G.,
  {Riggio}, A.: {Timing of the accreting millisecond pulsar XTE J1814-338}.
\newblock \mnras \textbf{375}(3), 971--976 (2007).
\newblock \doi{10.1111/j.1365-2966.2006.11359.x}

\bibitem{Papitto2009}
{Papitto}, A., {Di Salvo}, T., {D'A{\`\i}}, A., {Iaria}, R., {Burderi}, L.,
  {Riggio}, A., {Menna}, M.T., {Robba}, N.R.: {XMM-Newton detects a
  relativistically broadened iron line in the spectrum of the ms X-ray pulsar
  SAX J1808.4-3658}.
\newblock \aap \textbf{493}(3), L39--L43 (2009).
\newblock \doi{10.1051/0004-6361:200811401}

\bibitem{Papitto2013b}
{Papitto}, A., {Ferrigno}, C., {Bozzo}, E., {Rea}, N., {Pavan}, L., {Burderi},
  L., {Burgay}, M., {Campana}, S., {di Salvo}, T., {Falanga}, M.,
  {Filipovi{\'c}}, M.D., {Freire}, P.C.C., {Hessels}, J.W.T., {Possenti}, A.,
  {Ransom}, S.M., {Riggio}, A., {Romano}, P., {Sarkissian}, J.M., {Stairs},
  I.H., {Stella}, L., {Torres}, D.F., {Wieringa}, M.H., {Wong}, G.F.: {Swings
  between rotation and accretion power in a binary millisecond pulsar}.
\newblock \nat \textbf{501}(7468), 517--520 (2013).
\newblock \doi{10.1038/nature12470}

\bibitem{Papitto2011c}
{Papitto}, A., {Riggio}, A., {Burderi}, L., {di Salvo}, T., {D'A{\'\i}}, A.,
  {Iaria}, R.: {Spin down during quiescence of the fastest known
  accretion-powered pulsar}.
\newblock \aap \textbf{528}, A55 (2011).
\newblock \doi{10.1051/0004-6361/201014837}

\bibitem{Papitto2010}
{Papitto}, A., {Riggio}, A., {di Salvo}, T., {Burderi}, L., {D'A{\`\i}}, A.,
  {Iaria}, R., {Bozzo}, E., {Menna}, M.T.: {The X-ray spectrum of the newly
  discovered accreting millisecond pulsar IGR J17511-3057}.
\newblock \mnras \textbf{407}(4), 2575--2588 (2010).
\newblock \doi{10.1111/j.1365-2966.2010.17090.x}

\bibitem{Papitto2014}
{Papitto}, A., {Torres}, D.F., {Rea}, N., {Tauris}, T.M.: {Spin frequency
  distributions of binary millisecond pulsars}.
\newblock \aap \textbf{566}, A64 (2014).
\newblock \doi{10.1051/0004-6361/201321724}

\bibitem{2017ApJ...851L..34P}
{Parfrey}, K., {Tchekhovskoy}, A.: {General-relativistic Simulations of Four
  States of Accretion onto Millisecond Pulsars}.
\newblock \apjl \textbf{851}(2), L34 (2017).
\newblock \doi{10.3847/2041-8213/aa9c85}

\bibitem{Patruno2010}
{Patruno}, A.: {The Accreting Millisecond X-ray Pulsar IGR J00291+5934:
  Evidence for a Long Timescale Spin Evolution}.
\newblock \apj \textbf{722}(1), 909--918 (2010).
\newblock \doi{10.1088/0004-637X/722/1/909}

\bibitem{Patruno2012a}
{Patruno}, A.: {Evidence of Fast Magnetic Field Evolution in an Accreting
  Millisecond Pulsar}.
\newblock \apjl \textbf{753}(1), L12 (2012).
\newblock \doi{10.1088/2041-8205/753/1/L12}

\bibitem{Patruno2017}
{Patruno}, A.: {The Slow Orbital Evolution of the Accreting Millisecond Pulsar
  IGR J0029+5934}.
\newblock \apj \textbf{839}(1), 51 (2017).
\newblock \doi{10.3847/1538-4357/aa6986}

\bibitem{Patruno2009}
{Patruno}, A., {Altamirano}, D., {Hessels}, J.W.T., {Casella}, P., {Wijnands},
  R., {van der Klis}, M.: {Phase-Coherent Timing of the Accreting Millisecond
  Pulsar SAX J1748.9-2021}.
\newblock \apj \textbf{690}(2), 1856--1865 (2009).
\newblock \doi{10.1088/0004-637X/690/2/1856}

\bibitem{Patruno2010c}
{Patruno}, A., {Altamirano}, D., {Messenger}, C.: {The long-term evolution of
  the accreting millisecond X-ray pulsar SwiftJ1756.9-2508}.
\newblock \mnras \textbf{403}(3), 1426--1432 (2010).
\newblock \doi{10.1111/j.1365-2966.2010.16202.x}

\bibitem{Patruno2012b}
{Patruno}, A., {Bult}, P., {Gopakumar}, A., {Hartman}, J.M., {Wijnands}, R.,
  {van der Klis}, M., {Chakrabarty}, D.: {Accelerated Orbital Expansion and
  Secular Spin-down of the Accreting Millisecond Pulsar SAX J1808.4-3658}.
\newblock \apjl \textbf{746}(2), L27 (2012).
\newblock \doi{10.1088/2041-8205/746/2/L27}

\bibitem{Patruno2017b}
{Patruno}, A., {Haskell}, B., {Andersson}, N.: {The Spin Distribution of
  Fast-spinning Neutron Stars in Low-mass X-Ray Binaries: Evidence for Two
  Subpopulations}.
\newblock \apj \textbf{850}(1), 106 (2017).
\newblock \doi{10.3847/1538-4357/aa927a}

\bibitem{Patruno2016}
{Patruno}, A., {Maitra}, D., {Curran}, P.A., {D'Angelo}, C., {Fridriksson},
  J.K., {Russell}, D.M., {Middleton}, M., {Wijnand s}, R.: {The Reflares and
  Outburst Evolution in the Accreting Millisecond Pulsar SAX J1808.4-3658: A
  Disk Truncated Near Co-Rotation?}
\newblock \apj \textbf{817}(2), 100 (2016).
\newblock \doi{10.3847/0004-637X/817/2/100}

\bibitem{Patruno2012}
{Patruno}, A., {Watts}, A.L.: {Accreting Millisecond X-Ray Pulsars}.
\newblock arXiv e-prints arXiv:1206.2727 (2012)

\bibitem{Patruno2009d}
{Patruno}, A., {Wijnands}, R., {van der Klis}, M.: {An Alternative
  Interpretation of the Timing Noise in Accreting Millisecond Pulsars}.
\newblock \apjl \textbf{698}(1), L60--L63 (2009).
\newblock \doi{10.1088/0004-637X/698/1/L60}

\bibitem{Pintore2016}
{Pintore}, F., {Sanna}, A., {Di Salvo}, T., {Del Santo}, M., {Riggio}, A.,
  {D'A{\`\i}}, A., {Burderi}, L., {Scarano}, F., {Iaria}, R.: {Broad-band
  spectral analysis of the accreting millisecond X-ray pulsar SAX
  J1748.9-2021}.
\newblock \mnras \textbf{457}(3), 2988--2998 (2016).
\newblock \doi{10.1093/mnras/stw176}

\bibitem{Poutanen2006}
{Poutanen}, J.: {Accretion-powered millisecond pulsars}.
\newblock Advances in Space Research \textbf{38}(12), 2697--2703 (2006).
\newblock \doi{10.1016/j.asr.2006.04.025}

\bibitem{Psaltis2008}
{Psaltis}, D.: {Probes and Tests of Strong-Field Gravity with Observations in
  the Electromagnetic Spectrum}.
\newblock Living Reviews in Relativity \textbf{11}(1), 9 (2008).
\newblock \doi{10.12942/lrr-2008-9}

\bibitem{Rappaport2004}
{Rappaport}, S.A., {Fregeau}, J.M., {Spruit}, H.: {Accretion onto Fast X-Ray
  Pulsars}.
\newblock \apj \textbf{606}(1), 436--443 (2004).
\newblock \doi{10.1086/382863}

\bibitem{Riggio2011b}
{Riggio}, A., {Burderi}, L., {di Salvo}, T., {Papitto}, A., {D'A{\`\i}}, A.,
  {Iaria}, R., {Menna}, M.T.: {Secular spin-down of the AMP <ASTROBJ>XTE
  J1751-305</ASTROBJ>}.
\newblock \aap \textbf{531}, A140 (2011).
\newblock \doi{10.1051/0004-6361/201014883}

\bibitem{Riggio2008}
{Riggio}, A., {Di Salvo}, T., {Burderi}, L., {Menna}, M.T., {Papitto}, A.,
  {Iaria}, R., {Lavagetto}, G.: {Spin-up and Phase Fluctuations in the Timing
  of the Accreting Millisecond Pulsar XTE J1807-294}.
\newblock \apj \textbf{678}(2), 1273--1278 (2008).
\newblock \doi{10.1086/533578}

\bibitem{Riggio2011}
{Riggio}, A., {Papitto}, A., {Burderi}, L., {di Salvo}, T., {Bachetti}, M.,
  {Iaria}, R., {D'A{\`\i}}, A., {Menna}, M.T.: {Timing of the accreting
  millisecond pulsar IGR J17511-3057}.
\newblock \aap \textbf{526}, A95 (2011).
\newblock \doi{10.1051/0004-6361/201014322}

\bibitem{Riggio2020}
{Riggio}, A.e.a.: {IGR J17511-3057 in prep.}  (2020)

\bibitem{Romanova2005}
{Romanova}, M.M., {Ustyugova}, G.V., {Koldoba}, A.V., {Lovelace}, R.V.E.:
  {Propeller-driven Outflows and Disk Oscillations}.
\newblock \apjl \textbf{635}(2), L165--L168 (2005).
\newblock \doi{10.1086/499560}

\bibitem{Sanna2018a}
{Sanna}, A., {Bahramian}, A., {Bozzo}, E., {Heinke}, C., {Altamirano}, D.,
  {Wijnands}, R., {Degenaar}, N., {Maccarone}, T., {Riggio}, A., {Di Salvo},
  T., {Iaria}, R., {Burgay}, M., {Possenti}, A., {Ferrigno}, C., {Papitto}, A.,
  {Sivakoff}, G.R., {D'Amico}, N., {Burderi}, L.: {Discovery of 105 Hz coherent
  pulsations in the ultracompact binary IGR J16597-3704}.
\newblock \aap \textbf{610}, L2 (2018).
\newblock \doi{10.1051/0004-6361/201732262}

\bibitem{Sanna2018b}
{Sanna}, A., {Bozzo}, E., {Papitto}, A., {Riggio}, A., {Ferrigno}, C., {Di
  Salvo}, T., {Iaria}, R., {Mazzola}, S.M., {D'Amico}, N., {Burderi}, L.:
  {XMM-Newton detection of the 2.1 ms coherent pulsations from IGR
  J17379-3747}.
\newblock \aap \textbf{616}, L17 (2018).
\newblock \doi{10.1051/0004-6361/201833205}

\bibitem{Sanna2020c}
{Sanna}, A., {Burderi}, L., {Gendreau}, K.C., {Di Salvo}, T., {Ray}, P.S.,
  {Riggio}, A., {Gambino}, A.F., {Iaria}, R., {Piga}, L., {Malacaria}, C.,
  {Jaisawal}, G.K.: {Timing of the accreting millisecond pulsar IGR
  J17591-2342: evidence of spin-down during accretion}.
\newblock arXiv e-prints arXiv:2003.05069 (2020)

\bibitem{Sanna2017c}
{Sanna}, A., {Di Salvo}, T., {Burderi}, L., {Riggio}, A., {Pintore}, F.,
  {Gambino}, A.F., {Iaria}, R., {Tailo}, M., {Scarano}, F., {Papitto}, A.: {On
  the timing properties of SAX J1808.4-3658 during its 2015 outburst}.
\newblock \mnras \textbf{471}(1), 463--477 (2017).
\newblock \doi{10.1093/mnras/stx1588}

\bibitem{Sanna2018c}
{Sanna}, A., {Ferrigno}, C., {Ray}, P.S., {Ducci}, L., {Jaisawal}, G.K.,
  {Enoto}, T., {Bozzo}, E., {Altamirano}, D., {Di Salvo}, T., {Strohmayer},
  T.E., {Papitto}, A., {Riggio}, A., {Burderi}, L., {Bult}, P.M., {Bogdanov},
  S., {Gambino}, A.F., {Marino}, A., {Iaria}, R., {Arzoumanian}, Z.,
  {Chakrabarty}, D., {Gendreau}, K.C., {Guillot}, S., {Markwardt}, C., {Wolff},
  M.T.: {NuSTAR and NICER reveal IGR J17591-2342 as a new accreting millisecond
  X-ray pulsar}.
\newblock \aap \textbf{617}, L8 (2018).
\newblock \doi{10.1051/0004-6361/201834160}

\bibitem{Sanna2017a}
{Sanna}, A., {Papitto}, A., {Burderi}, L., {Bozzo}, E., {Riggio}, A., {Di
  Salvo}, T., {Ferrigno}, C., {Rea}, N., {Iaria}, R.: {Discovery of a new
  accreting millisecond X-ray pulsar in the globular cluster NGC 2808}.
\newblock \aap \textbf{598}, A34 (2017).
\newblock \doi{10.1051/0004-6361/201629406}

\bibitem{Sanna2017d}
{Sanna}, A., {Pintore}, F., {Bozzo}, E., {Ferrigno}, C., {Papitto}, A.,
  {Riggio}, A., {Di Salvo}, T., {Iaria}, R., {D'A{\`\i}}, A., {Egron}, E.,
  {Burderi}, L.: {Spectral and timing properties of IGR J00291+5934 during its
  2015 outburst}.
\newblock \mnras \textbf{466}(3), 2910--2917 (2017).
\newblock \doi{10.1093/mnras/stw3332}

\bibitem{Sanna2019}
{Sanna}, A., {Pintore}, F., {Riggio}, A., {Burderi}, L., {Gambino}, A.F.,
  {Gendreau}, K.C., {Arzoumanian}, Z., {Bult}, P.M., {di Salvo}, T., {Iaria},
  R., {Ferrigno}, C., {Bozzo}, E., {Papitto}, A.: {NICER and SWIFT/XRT detect a
  new outburst of the accreting millisecond X-ray pulsar SWIFT J1756.9-2508.}
\newblock The Astronomer's Telegram, \textbf{12882}, 1 (2019)

\bibitem{Sanna2018d}
{Sanna}, A., {Pintore}, F., {Riggio}, A., {Mazzola}, S.M., {Bozzo}, E., {Di
  Salvo}, T., {Ferrigno}, C., {Gambino}, A.F., {Papitto}, A., {Iaria}, R.,
  {Burderi}, L.: {SWIFT J1756.9-2508: spectral and timing properties of its
  2018 outburst}.
\newblock \mnras \textbf{481}(2), 1658--1666 (2018).
\newblock \doi{10.1093/mnras/sty2316}

\bibitem{Sanna2017b}
{Sanna}, A., {Riggio}, A., {Burderi}, L., {Pintore}, F., {Di Salvo}, T.,
  {D'A{\`\i}}, A., {Bozzo}, E., {Esposito}, P., {Segreto}, A., {Scarano}, F.,
  {Iaria}, R., {Gambino}, A.F.: {Study of the accretion torque during the 2014
  outburst of the X-ray pulsar GRO J1744-28}.
\newblock \mnras \textbf{469}(1), 2--12 (2017).
\newblock \doi{10.1093/mnras/stx635}

\bibitem{Sanna2020}
{Sanna}, A.e.a.: {SAX J1748.9-2021 in prep.}  (2020)

\bibitem{Sanna2020b}
{Sanna}, A.e.a.: {SAX J1808.4-3658 in prep.}  (2020)

\bibitem{Spitkovsky2006}
{Spitkovsky}, A.: {Time-dependent Force-free Pulsar Magnetospheres:
  Axisymmetric and Oblique Rotators}.
\newblock \apjl \textbf{648}(1), L51--L54 (2006).
\newblock \doi{10.1086/507518}

\bibitem{Strohmayer2017}
{Strohmayer}, T., {Keek}, L.: {IGR J17062-6143 Is an Accreting Millisecond
  X-Ray Pulsar}.
\newblock \apjl \textbf{836}(2), L23 (2017).
\newblock \doi{10.3847/2041-8213/aa5e51}

\bibitem{Tailo2018}
{Tailo}, M., {D'Antona}, F., {Burderi}, L., {Ventura}, P., {di Salvo}, T.,
  {Sanna}, A., {Papitto}, A., {Riggio}, A., {Maselli}, A.: {Evolutionary paths
  of binaries with a neutron star - I. The case of SAX J1808.4 - 3658}.
\newblock \mnras \textbf{479}(1), 817--828 (2018).
\newblock \doi{10.1093/mnras/sty1637}

\bibitem{Tauris2001}
{Tauris}, T.M., {Konar}, S.: {Torque decay in the pulsar (P,dot \{P\}) diagram.
  Effects of crustal ohmic dissipation and alignment}.
\newblock \aap \textbf{376}, 543--552 (2001).
\newblock \doi{10.1051/0004-6361:20010988}

\bibitem{vanParadijs1994}
{van Paradijs}, J., {McClintock}, J.E.: {Absolute visual magnitudes of low-mass
  X-ray binaries.}
\newblock \aap \textbf{290}, 133--136 (1994)

\bibitem{Wang2017}
{Wang}, L., {Steeghs}, D., {Casares}, J., {Charles}, P.A., {Mu{\~n}oz-Darias},
  T., {Marsh}, T.R., {Hynes}, R.I., {O'Brien}, K.: {System mass constraints for
  the accreting millisecond pulsar XTE J1814-338 using Bowen fluorescence}.
\newblock \mnras \textbf{466}(2), 2261--2271 (2017).
\newblock \doi{10.1093/mnras/stw3312}

\bibitem{Wang87}
{Wang}, Y.M.: {Disc accretion by magnetized neutron stars - A reassessment of
  the torque}.
\newblock \aap \textbf{183}, 257--264 (1987)

\bibitem{Wang2013}
{Wang}, Z., {Breton}, R.P., {Heinke}, C.O., {Deloye}, C.J., {Zhong}, J.:
  {Multiband Studies of the Optical Periodic Modulation in the X-Ray Binary SAX
  J1808.4-3658 during Its Quiescence and 2008 Outburst}.
\newblock \apj \textbf{765}(2), 151 (2013).
\newblock \doi{10.1088/0004-637X/765/2/151}

\bibitem{Wijnands1998}
{Wijnands}, R., {van der Klis}, M.: {A millisecond pulsar in an X-ray binary
  system}.
\newblock \nat \textbf{394}(6691), 344--346 (1998).
\newblock \doi{10.1038/28557}

\bibitem{Wijnands1999}
{Wijnands}, R., {van der Klis}, M.: {The Broadband Power Spectra of X-Ray
  Binaries}.
\newblock \apj \textbf{514}(2), 939--944 (1999).
\newblock \doi{10.1086/306993}

\bibitem{Will2006}
{Will}, C.M.: {The Confrontation between General Relativity and Experiment: A
  Centenary Perspective}.
\newblock Progress of Theoretical Physics Supplement \textbf{163}, 146--162
  (2006).
\newblock \doi{10.1143/PTPS.163.146}

\bibitem{Ziolkowski2018}
{Zi{\'o}{\l}kowski}, J., {Zdziarski}, A.A.: {Non-conservative mass transfer in
  stellar evolution and the case of V404 Cyg/GS 2023+338}.
\newblock \mnras \textbf{480}(2), 1580--1586 (2018).
\newblock \doi{10.1093/mnras/sty1948}

\end{thebibliography}

\end{document}